\documentclass[superscriptaddress,twocolumn,showpacs,preprintnumbers,amsmath,amssymb,floatfix,nofootinbib]{revtex4-1}
\usepackage{color}   %option added by cyw eg \pagecolor{blue} \color{red}
\usepackage{graphicx}% Include figure files
\usepackage{dcolumn}% Align table columns on decimal point
\usepackage{float}

\begin{document}

\def\bb    #1{\hbox{\boldmath${#1}$}}

\title{ Signature of the  Fragmentation of a Color Flux Tube} 

\author{Cheuk-Yin Wong}
\email{wongc@ornl.gov}
\affiliation{Physics Division, Oak Ridge National Laboratory,
Oak Ridge, Tennessee 37831, USA}

\begin{abstract}
The production of quark-antiquark pairs along a color flux tube
precedes the fragmentation of the tube.  Because of local conservation
laws, the production of a $q$-$\bar q$ pair will lead to correlations
of adjacently produced mesons (mostly pions).  Adjacently produced
mesons however can be signalled by their rapidity difference $\Delta
y$ falling within the window of $|\Delta y | $$\lesssim$$ 1/(dN/dy)$,
on account of the space-time-rapidity ordering of produced mesons in a
flux-tube fragmentation.  Therefore, the local conservation laws of
momentum, charge, and flavor will lead to a suppression of the angular
correlation function $dN/(d\Delta \phi\, d\Delta y)$ for two mesons
with opposite charges or strangeness on the near side at $(\Delta
\phi, \Delta y)$$ \sim$0, but an enhanced correlation on the
back-to-back, away side at $\Delta \phi$$\sim$$ \pi$, within the
window of $|\Delta y |$$\lesssim$$ 1/(dN/dy)$.  These properties can
be used as signatures for the fragmentation of a color flux tube.  The
gross features of the signature of flux-tube fragmentation for two
oppositely charged mesons are qualitatively consistent with the STAR
and NA61/SHINE angular correlation data for two hadrons with opposite
charges in the low-$p_T$ region in high-energy $pp$ collisions.

\end{abstract}


\maketitle
\section{Introduction}
\label{Introduction}

In the production of particles in the central rapidity region in
high-energy hadron-hadron collisions and $e^+ e^-$ annihilations, the
low-$p_T$ part of the spectra falls within the realm of soft
nonperturbative QCD physics and is usually considered to arise from
the fragmentation of a color flux tube (or its idealization as a QCD
string)
\cite{Sch51,Sch62,Nam70,Bjo73,Cas74,Art74,And79,And83,And83a,Art84,Sjo86,Wan88,Pav91,Won91a,Won91b,Gat92,Won95,Feo08,Won94}.
The high-$p_T$ part in hadron-hadron collisions is considered to arise
from a relativistic hard scattering of partons and subsequent parton
showering
\cite{Bla74,Ang78,Ang79,Fey78,Owe78,Duk84,Sj87,Sjo14,UA188,Wan91,Won94,Arl10,Rak13,Won12,Won13a,Won13}.
In the case of high-energy $e^+e^-$ annihilation, the annihilation
leads to the production of high-energy quark and antiquark partons
which are then subject to perturbative QCD parton showering processes
to lead to the production of hadrons.

Recently, it was found that the hadron $p_T$ spectra spanning over 14
decades of magnitude from about 0.5 GeV/c to the highest $p_T$ at
central rapidity in $pp$ collisions at LHC energies can be adequately
described by a Tsallis distribution with only three degrees of freedom
\cite{Won12,Won13,Won13a,Won14EPJ,CTWW14,T1,Won15}, in a form
phenomenologically equivalent to the quasi-power law introduced by
Hagedorn \cite{H} and others \cite{Michael} for relativistic hard
scattering.  The simplicity of the $p_T$ spectrum suggests that a
single mechanism dominates over the domain with $p_T$$>$0.5 GeV/c at
central rapidity in these high-energy collisions.  As the high-$p_T$
region is known to arise from the relativistic hard-scattering process
\cite{Bla74,Ang78,Ang79,Fey78,Owe78,Duk84,Sj87,Sjo14,UA188,Wan91,Won94,Arl10,Rak13,Won12,Won13,Won13a},
one is led to the suggestion that the hard-scattering process
dominates over the domain with $p_T$$>$0.5 GeV/c in these high-energy $pp$
collisions.  Additional experimental evidences have been uncovered to
support such a suggestion \cite{Won15}.

The dominance of the hard-scattering process does not imply the
absence of the soft flux-tube fragmentation process.  It only
stipulates that contributions from the hard-scattering process
increase with increasing collision energies and the fraction of the
contributions from the flux-tube fragmentation process becomes smaller
in comparison, as pointed out earlier in \cite{UA188,Wan91}.  As a
consequence, there will be a transverse momentum boundary
$p_{Tb}$ which separates the lower domain of flux-tube fragmentation
 from the higher domain of hard-scattering dominance. 

As the $pp$ collision energy decreases, the role of the flux-tube
fragmentation and hard scattering will be reversed, with an increase
in the fraction of contributions from the flux-tube fragmentation and
a shift of the transverse momentum boundary $p_{Tb}$ to greater $p_T$
values.  It is of interest to see how the two processes interplay and
how the boundary function $p_{Tb}(\sqrt{s_{NN}}) $ between the two
processes depends on the collision energy $\sqrt{s_{NN}}$.  In
addition to being an intrinsic physical property of the $pp$ collision
process, the boundary function $p_{Tb} (\sqrt{s_{NN}})$ separating the
two processes in $pp$ collisions may have implications on the early
evolution dynamics, the thermalization of the produced medium, the quenching
of jets, and the formation of the quark-gluon plasma, in high-energy
nucleus-nucleus collisions.  It is therefore desirable to search for
ways to discriminate the process of flux-tube fragmentation from the
process of the hard scattering in $pp$ collisions so that they can be
separated out and the boundary $p_{Tb}$ mapped out as a
function of the collision energy.  We need well-defined signatures for
the flux-tube fragmentation and the hard-scattering processes.

Two-hadron $\Delta \phi$-$\Delta \eta$ angular correlations have been
previously first suggested and used by the STAR Collaboration to separate
phenomenologically the `soft' and `hard' components in nucleon-nucleon
and nucleus-nucleus collisions \cite{STAR06twopar,Por05,Tra11,Ray11,TraKet11}.
The signature for the hard component, represented by the production of
two back-to-back jets (minijets) in the hard-scattering process, is
well known \cite{Ang79,Fey78,Owe78,Duk84,Sj87,Sjo14,Rak13}, as indicated for
example by STAR $\Delta\phi$-$\Delta \eta$ correlation data for two
oppositely charged hadrons with $p_T$$>$0.5 GeV/c in $pp$ collisions
at $\sqrt{s_{NN}}$=200 GeV shown in Fig.\ 1(b)
\cite{STAR06twopar,Por05,Tra11,Ray11,TraKet11}.  It consists of (i) a near-side
cluster of particles at $(\Delta\phi, \Delta\eta)$$\sim$0 for one jet,
and (ii) an away-side ridge at $\Delta\phi$$ \sim $$\pi$ along $
\Delta\eta$ for the other jet.
\begin{figure}[h]
\label{fig1}
\includegraphics[width = 230 pt, height = 115 pt]{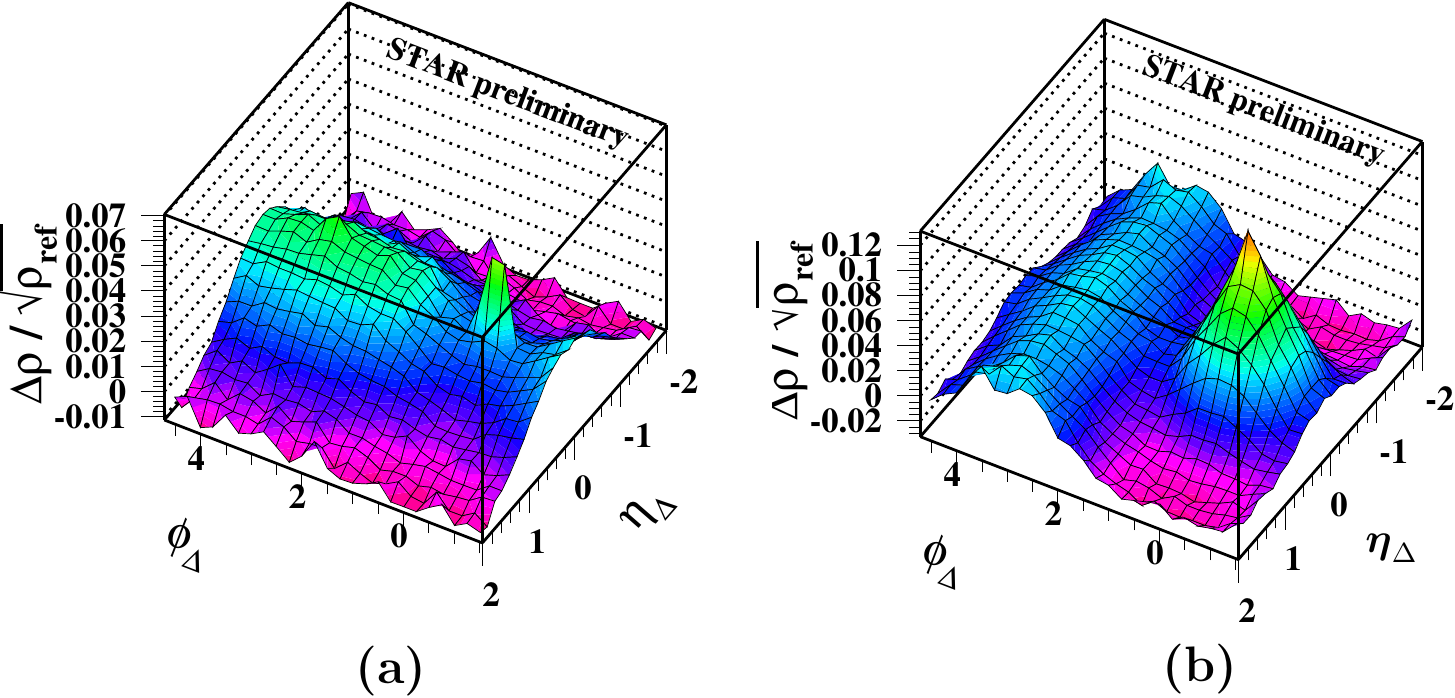} 
\caption{\label{etaphi} (Color online) The angular correlation data 
$\Delta \rho/\sqrt{\rho_{\rm ref}}$ as a function 
of the
  azimuthal angle difference $\phi_\Delta$=$\Delta \phi$ and
  pseudorapidity difference $\eta_\Delta$=$\Delta \eta$ of two
  oppositely charged hadrons in $pp$ collisions at $\sqrt{s_{NN}}$=200
  GeV (from Fig. 3 of \cite{Por05} of the STAR Collaboration).  Fig. 1(a) is for hadrons in the
  domain with $p_T$$<$0.5 GeV/c.   Fig.\ 1(b) is for hadrons in the domain with
  $p_T$$>$0.5 GeV/c }
\end{figure} 

On the other hand, the `soft' component,  defined phenomenologically by the STAR Collaboration
to be the mechanism that dominates the production process in the domain with
$p_T$$<$0.5 GeV/c for $pp$ collisions at $\sqrt{s_{NN}}$=200 GeV,  
was  associated with the
$(\Delta\phi, \Delta\eta)$ correlation for two oppositely charged
hadrons as shown in Fig.\ 1(a) \cite{STAR06twopar,Por05,Tra11,Ray11,TraKet11}.  
 The small narrow peak at $(\Delta \phi,\Delta \eta)\sim 0$  is an experimental   detector artifact.  After removing this artifact, 
the correlation is suppressed at
$(\Delta \phi, \Delta \eta)$$\sim$0 but enhanced at $(\Delta
\phi$$\sim $$\pi, \Delta \eta$$\sim$0).  The two-hadron correlation of
the soft component was also taken phenomenologically 
as a one-dimensional Gaussian distribution at $\Delta
\eta$$\sim$0  independent of $\Delta \phi$
\cite{STAR06twopar,Por05,Tra11,Ray11,TraKet11}.  A rigorous proof of the
microscopic connection between the two-hadron $\Delta\phi$-$\Delta
\eta$ correlation and the signature of flux-tube fragmentation is
still lacking.  The other suggested signature of a one-dimensional Gaussian
at $\Delta \eta$$ \sim$0 also needs to be reexamined carefully.
Starting from the constituent level, the microscopic connection of
$dN/(d \Delta \phi)( d\Delta \eta)$ in a flux-tube fragmentation will
be the subject of the present investigation.

The search for the signature of the fragmentation of the flux tube is
also useful for a better understanding of the multiparticle dynamics of
the fragmentation process.  Experimental verification of the signature
will pave the way for further investigations to find out how a leading
quark and an antiquark pair can produce a chain of hadrons that are
ordered in space-time and rapidities through the nonperturbative QCD
processes.  Of particular interest is the possibility of an
event-by-event exclusive measurement in which all of the momenta of
the produced particles in a reaction event are measured, matched, and
correlated, as in a jigsaw puzzle, to see how the space-time-rapidity
ordering may allow the event reconstruction of the chain of hadrons at
the moment of flux-tube fragmentation.

To search for the signature, we envisage that in a flux-tube
fragmentation in hadron-hadron collisions or $e^+e^-$ annihilations, a
flux tube is initially formed between the leading quark and antiquark
(or diquark in the case of a nucleon-nucleon collision) when they pull
apart from each other at high energies.  The vacuum is so polarized
that ordered pairs of quarks and antiquarks are produced inside the
tube via the Schwinger pair-production mechanism or the QED2
inside-outside cascade mechanism
\cite{Sch51,Sch62,Bjo73,Cas74,Wan88,Pav91,Won91a,Won91b,Gat92,Won94,Won95}.
The interaction of the produced quarks with antiquarks produced in
adjacent vertices leads to the production of mesons and the
fragmentation of the flux tube.

The production of a quark-antiquark pair needs to obey local
conservation laws.  These conservation laws will impose constraints
and will lead to correlations of various quantities, for two
longitudinally adjacent produced mesons.  The occurrence of two
longitudinally adjacent produced mesons, on the other hand, is
signalled by their proximity in rapidity, on account of the
space-time-rapidity ordering of the produced mesons (see for example,
Chapter 6 of \cite{Won94}).  Hence, the local conservation laws will
lead to correlations of various quantities as a function of the
azimuthal angle difference $\Delta \phi$ and rapidity difference
$\Delta y$ (or approximately, the pseudorapdity difference $\Delta
\eta$) for a pair of produced mesons in a flux-tube fragmentation.
These angular correlations can be used as the signature of the flux
tube fragmentation in high-energy nucleon-nucleon 
collisions and $e^+$-$e^-$ annihilations.

In this paper, we restrict our attention to the central rapidity
region.  In Section II, for a system of four produced constituents we
single out the relevant degrees of freedom to describe the flux-tube
fragmentation process.  In Section III, we discuss the momentum
distribution of two nonadjacent mesons when the quark and antiquark
constituents in the detected mesons are produced without
correlations. In Section IV, we examine the transverse momentum
distribution of adjacent mesons in which the quark of one meson and
the antiquark of the other meson are produced at the same point,
subject to local conservation of momentum.  In Section V, we discuss
the rapidity distribution of produced mesons in a flux tube.  In
Section VI, we examine the charge correlation of two adjacent and
nonadjacent mesons for a system with two flavors.  In Section VII, we
consider the correlation of charge and strangeness for two adjacent
and nonadjacent mesons for a system with three flavors.  In Section
VIII, we present the two-particle correlation function for
measurements with different combinations of meson charges and
strangeness quantum numbers.  In Section IX, we present 
numerical examples of theoretical correlation functions
of $dN/d\Delta \phi d\Delta \eta$ for different charge and strangeness configurations
 to provide   signatures for  flux-tube fragmentation.
In Section X,
we compare the
theoretical signature of the correlation of two  oppositely
charged hadrons with STAR and NA61/SHINE two-hadron correlation data for $pp$
collisions
\cite{STAR06twopar,Por05,Tra11,Ray11,TraKet11,Mak15,Gaz15,Lar15,Ser15}.  In
Section XI, we present our conclusions and discussions.

\section{ production of Two Mesons in flux-tube Fragmentation }

Our objective is to search for the signature of flux-tube
fragmentation by studying the correlations between two mesons produced
in a flux tube.  Starting with the microscopic constituent momenta, it
is necessary to enumerate the relevant degrees of freedom before we
can examine the correlations of the observed composite mesons.

\begin{figure}[h]
  \centering
\includegraphics[scale=0.40]{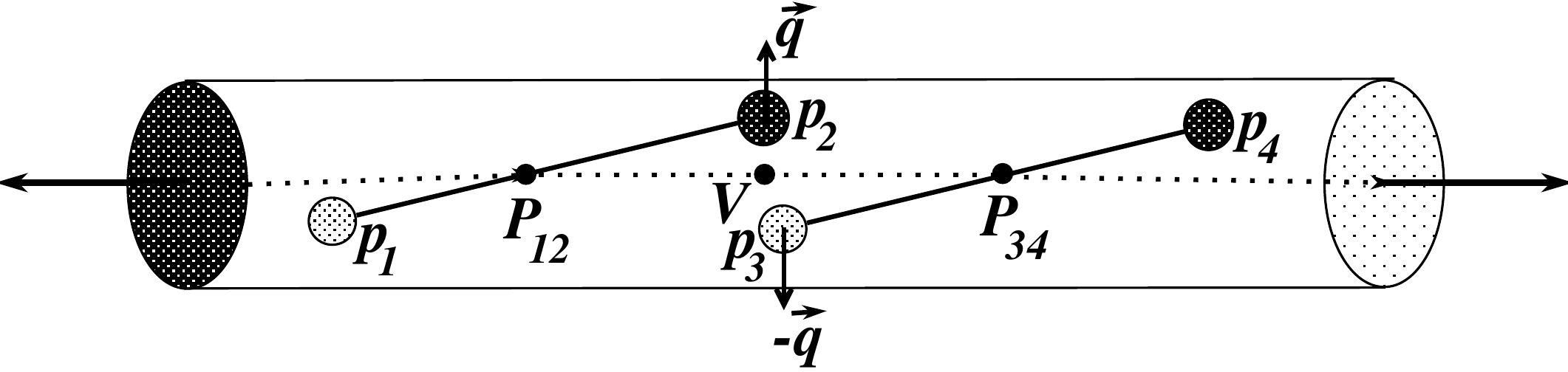}
\caption{ Schematic depiction of the production of quarks and
  antiquarks in two adjacent mesons $P_{12}$ and $P_{34}$ as the
  leading quark is pulling to the left and the leading antiquark is
  pulling to the right, with the production of $p_1$(antiquark),
  $p_2$(quark), $p_3$(antiquark) and $p_4$(quark) along the flux tube.
  Subsequently $p_1$ interacts with $p_2$ to form meson $P_{12}$,
  while $p_3$ interacts with $p_4$ to form meson $P_{34}$.  }
\label{fig2}
\end{figure}

We consider the production of two adjacent mesons in a flux tube at
the moment of its fragmentation, as depicted schematically in
Fig.\ \ref{fig2} where the leading quark pulls apart in one direction
while the other leading antiquark pulls apart in the other direction
at high energies.  As the leading quark and antiquark pull apart, a
flux tube is formed and the vacuum in the tube is polarized.  The
flux-tube fragmentation process is initiated by the production of many
ordered $q$-$\bar q$ pairs along the flux tube by the Schwinger
particle-production mechanism or the inside-outside cascade mechanism,
as discussed in Refs.\ 
\cite{Sch51,Sch62,Bjo73,Cas74,Art74,And79,And83,And83a,Sjo14,Art84,Sjo86,Wan88,Pav91,Won91a,Won91b,Gat92,Won94,Feo08,Won95}.
The production of these $q$-$\bar q$ pairs takes place locally at
spatial points along the tube.

We focus our attention in a small section in the central rapidity
region in the center-of-mass system and consider the pair $p_2$(quark)
and $p_3$(antiquark) produced at the vertex $V$ in Fig.\ \ref{fig2}.
We shall use the particle label also to represent the particle
momentum.  After $p_2$ and $p_3$ are produced, $ p_2$ interacts with
the adjacent antiparticle $p_1$ with a confining interaction, leading
to the formation of the meson $P_{12}$.  Similarly, $ p_3$ interacts
with the adjacent $p_4$ to form the adjacent meson $P_{34}$.  In the
leading order of the confining interaction, the constituents of meson
$P_{12}$ and the constituents of the adjacent meson $P_{34}$ do not
interact\footnote{There are however higher-order residual interactions
  between mesons, as discussed for example by Peshkin and Bhanot
  \cite{Pes79} and in Refs.\  \cite{Ynd02,Won04}.} because the linearly confining
interactions are screened by their partners.  The formation of the
nearly noninteracting mesons $ P_{12}$ and $ P_{34}$ leads to the
fragmentation of the flux tube into mesons along the longitudinal
direction, populating the longitudinal momentum space in the form of a
rapidity plateau.

The pair-wise two-body interaction between $p_1$ and $p_2$ as well as
between $p_3$ and $p_4$ along the flux tube makes it simple to study
the many-body problem of quarks and antiquarks in two-body basis
\cite{Won01,Won04}.  We can use Dirac's constraint dynamics
\cite{dirac,cra82,cra88,saz89,cra92,cra96,Won01} to separate out the
coordinates of a pair of particles into their center-of-mass and
relative coordinates in a relativistically consistent manner.

 For the pair of interacting particles with 4-momenta $p_1$ and $p_2$
 and rest masses $m_1$ and $m_2$, we construct the total momentum
 $P_{12}$ and the relative momentum $p_{12}$,
\begin{subequations}
\begin{equation}
P_{12}=p_{1}+p_{2},
\end{equation}
\begin{equation}
p_{12}=
\frac{p_2\cdot P_{12}}{P_{12}^2} p_1 
-\frac{p_1\cdot P_{12}}{P_{12}^2} p_2,  \label{qrel}
\end{equation}
\end{subequations}
where 
\begin{subequations}
\begin{eqnarray}
{\frac{p_{1}\cdot P_{12}}{{P_{12} ^{2}}}}&&=\frac{
P_{12} ^{2}+p_{1}^{2}-p_{2}^{2}}{2{P_{12}^{2}}}  \\
\hspace*{-2.0cm}{\rm and~~~~~~~~~~~~~~~~~~~~~} {\frac{p_{2}\cdot P_{12}}{{P_{12}^{2}}}}&&=\frac{
P_{12} ^{2}+p_{2}^{2}-p_{1}^{2}}{2{P_{12}^{2}}}  \label{ep12}
\end{eqnarray}
\end{subequations}
are the projections of the momenta $p_1$ and $p_2$ along the direction
of the total momentum $P_{12}$.  The inverse transformation is
\begin{subequations}
\label{3}
\begin{eqnarray}
p _{1} &=&{\frac{p_{1}\cdot P_{12}}{{P_{12} ^{2}}}}P_{12} +p_{12},\\
p _{2} &=&{\frac{p_{2}\cdot P_{12}}{{P_{12} ^{2}}}}P_{12} - p_{12}.
\end{eqnarray}
\end{subequations}
Under the interaction $\Phi(x_{\perp 12})$ which depends only on the
time-transverse relative coordinate $x_{\perp 12}$ between $p_1$ at $x_1$ and
$p_2$ at $x_2$, namely,
\begin{eqnarray}
x_{\perp 12}= (x_1-x_2) -\frac{ (x_1 - x_2 )\cdot P_{12} }{P_{12}^2} P_{12},
\end{eqnarray}
with the property $x_{\perp 12} \cdot P_{12}=0$, the eigenvalue equation for the meson bound state with wave function $\psi(x_{\perp 12})$ in the relative coordinate is
\cite{dirac,cra82,cra88,saz89,cra92,cra96,Won01}
\begin{eqnarray}
[b^{2}(P_{12}^{2};m_{1}^{2},m_{2}^{2})+\bb p_{12}^{2}-\Phi
(x_{\perp 12 })]|\psi (x_{\perp 12}) \rangle=0,  \label{sep}
\end{eqnarray}
where 
\begin{equation}
\hspace*{-0.2cm}
b^{2}(P_{12}^{2},m_{1}^{2},m_{2}^{2})\!=\!\frac{
P_{12}^{4}\!-\!2P_{12}^{2}(m_{1}^{2}\!+\!m_{2}^{2})\!+\!(m_{1}^{2}\!-\!m_{2}^{2})^{2}
}{4P_{12}^{2}},
\end{equation}
\begin{eqnarray}
P_{12}^2=M_{12}^2,
\label{7}
\end{eqnarray}
and
\begin{eqnarray}
M_{12}=\sqrt{b^2 + m_1^2}+\sqrt{b^2+m_2^2}.
\end{eqnarray}
The solution of the bound state equation (\ref{sep}) then leads to the
meson wave function $\psi(x_{\perp 12})$ and the meson mass
$M_{12}$.\footnote{In practice, the quark and antiquark have
  spins, and one needs to use the two-body Dirac equation in place of
  Eq.\ (\ref{sep}) for the eigenvalue equation involving a
  color-Coulomb plus a linear confining interaction.  The numerical
  solution of the relativistic two-body Dirac equation then leads to
  the meson state with the meson mass $M_{12}$ and the proper
  meson spatial wave function, as described in Refs.\ \cite{Cra04,Cra09} and
  references cited therein.  A nonrelativistic model of a meson as
  a $q$-$\bar q$ bound state is given in Refs.\ \cite{Bar92,Won02}.}
Equations (\ref{sep}) and (\ref{7}) indicate that as a result of the
interaction $\Phi(x_{\perp 12})$ that exists between $p_1$ and $p_2$,
the momentum elements are related by
\begin{eqnarray}
&&dp_1 \delta (p_1^2 - m_1^2) dp_2 \delta  (p_2^2 - m_2^2) 
\nonumber\\
&&\!=\! dP_{12} \delta (P_{12}^2\!-\!M_{12}^2) dp_{12} \delta ( b^2\! -\! \bb p_{12}^2 \!-\!\Phi(x_{\perp 12}))\delta(p_{0,12}).~~
\end{eqnarray}

In the context of the fragmentation of the a flux tube, the produced
antiquark $p_1$ and the quark $p_2$ line up along the tube as in a
one-dimensional string at the initial moment of meson production, and
one can use the string approximation to separate out the longitudinal
and transverse degrees of freedom by approximating the $\bb p_{12}$ in
Eq.\ ({\ref{sep}) by $p_{z 12}$ and the confining potential coordinate
  $x_{\perp 12}$ to be longitudinal.  The longitudinal $p_{z 12}$ and
  the transverse degrees of freedom can then be separated.  The
  solution of the bound-state problem takes care of the $p_{z 12}$
  degrees of freedom.  Upon separating out and subsequently
  integrating over the variables of $P_{0, 12}$, $p_{0,12}$ and $p_{z,
    12}$, the remaining relevant degrees of freedom in the momentum
  element for particles $p_1$ and $p_2$ in the flux-tube fragmentation
  are
\begin{eqnarray}
\frac {dP_{z 12}}{E_{12}} d\bb P_{T12} d\bb p_{T 12}= dy_{12}d\bb P_{T12} d\bb p_{T 12} , 
\end{eqnarray}
where $y_{12}$ is the rapidity of the meson $P_{12}$, and the
subscript $T$ represents the transverse (two-dimensional) components.

Similarly, the particles $p_3$ and $p_4$ can be described by a
composite meson with total momentum $P_{34}$
\begin{subequations}
\label{11}
\begin{eqnarray}
P_{34}&&=p_{3}+p_{4},
\\
p_{34}&&=\frac{p_4 \cdot P_{34}}{P_{34}^{2}}   p_{3}
-\frac{p_3 \cdot P_{34}}{P_{34}^{2}}   p_{4},
\end{eqnarray}
\end{subequations}
with a relative momentum $p_{34}$ containing a bound state with the
meson mass $P_{34}^2=M_{34}^2$.  The inverse transformation is
\begin{subequations}
\label{12b}
\begin{eqnarray}
p _{3} &=&{\frac{p_{3}\cdot P_{34}}{{P_{34} ^{2}}}}P_{34} +p_{34},\\
p _{4} &=&{\frac{p_{4}\cdot P_{34}}{{P_{34} ^{2}}}}P_{34} - p_{34}.
\end{eqnarray}
\end{subequations}
The relevant degrees of freedom in the momentum element for particles
$p_3$ and $p_4$ in the flux-tube fragmentation are
\begin{eqnarray}
dy_{34}d\bb P_{T34} d\bb p_{T 34} .
\end{eqnarray}
Thus, for two mesons $P_{12}$ and $P_{34}$ in the problem of flux-tube
fragmentation, the probability distribution $dN$ is in general a
function of $dN(y_{12},y_{34}, \bb P_{T12}, \bb P_{T34}, \bb p_{T12},
\bb p_{T34})$.

\section{The Two-Particle Transverse  Distribution 
for nonadjacently Produced  Mesons} 

The process of $q$-$\bar q$ production for two adjacent mesons
$P_{12}$ and $P_{34}$ depicted in Fig.\ \ref{fig2} occurs in the
neighborhood of the vertex $V$.  We envisage that similar processes of
$q$-$\bar q$ production occur independently at many other vertices
along the flux tube, leading to the production of mesons populating
the longitudinal momentum space in the form of a rapidity plateau.

We can consider two of the produced mesons which can be related to
each other in two different ways.  The mesons $P_{12}$, and $P_{34}$
can be in a (longitudinally) adjacent state as in Fig.\ \ref{fig2}, in
which the quark $p_2$ of $P_{12}$ and the antiquark $p_3$ of $P_{34}$
are 
produced at the same spatial point  $V$ and are correlated.  The two mesons
$P_{12}$, and $P_{34}$ can be in a (longitudinally) nonadjacent state
as in Fig.\ \ref{fig3}, in which all quarks and the antiquarks of the
two mesons are independently produced at different vertices and are
not correlated with each other.  We shall often label the state of two
mesons by the superscript $X$ with $X$=$A$ for the adjacent state and
$X$=$N$ for the nonadjacent state.
 
\begin{figure}[h]
  \centering
\includegraphics[scale=0.40]{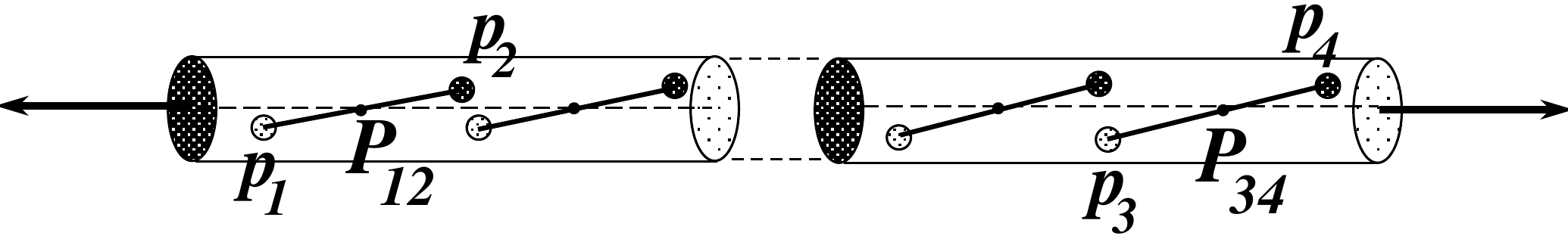}
\caption{ Schematic depiction of the production of quarks and antiquarks $p_1$, $p_2$, $p_3$, and 
 $p_4$ in two nonadjacent mesons
  $P_{12}$ and $P_{34}$, in a  flux-tube fragmentation.  The quark
  $p_2$ and the antiquark $p_3$ are not correlated because they are
  produced at different vertices.  }
\label{fig3}
\end{figure}

In the production of the four constituents $p_1,p_2,p_3,p_4$, we can
separate out the longitudinal and transverse degrees of freedom to
write the distribution as
\begin{eqnarray}
dN=dN_y dN_T.
\end{eqnarray}
 The probability for the two-meson state to be adjacent or
 nonadjacent depends mainly on the magnitude of their longitudinal
 rapidity differences $|\Delta y|$ in comparison with the inverse of
 the rapidity density, $1/(dN/dy)$, which we shall take up in Section
 V.  We examine in this section the transverse momentum distribution
 of nonadjacent meson, $dN_T^N$, and in the next section the the
 transverse momentum distribution of adjacent mesons, $dN_T^A$.

The quarks and antiquarks of the constituents of the mesons are
produced inside the flux tube, the wave functions need to obey the
boundary condition appropriate for the tube.  They acquire a
transverse momentum distribution governed by the geometry of the tube
with a standard deviation $\sigma_u$ related to the radius of
the tube $R_{\rm tube}$ by
\begin{eqnarray}
\sigma_u \sim \frac{\hbar }{R_{\rm tube}}.
\end{eqnarray}•
In addition, the quark and antiquark pair are produced in a Schwinger
pair-production mechanism, from which they acquire a transverse
momentum distribution with a standard deviation $\sigma_q$ given by
\cite{Wan88,Pav91,Won94,Won95}
\begin{eqnarray}
\sigma_q=\sqrt{\frac{\kappa}{\pi}},
\end{eqnarray}
where $\kappa$ is the strength of the linear color-electric field
between the polarizing leading quark and antiquark.  By folding the
transverse momentum distribution from the flux tube with the transverse momentum
distribution from the Schwinger mechanism, the transverse momentum 
distribution $dN_{Ti}$
of each constituent $\bb p_{Ti}$ can be represented by a
Gaussian distribution as
\begin{eqnarray}
\frac{dN_{Ti}}{d\bb p_{Ti}} 
&&=  \frac{1}{(\sqrt{2\pi} \sigma_u)^2} \int 
\frac{d\bb q~ \exp\{-\frac{\bb q^2}{2\sigma_q^2}\}}
{(\sqrt{2\pi} \sigma_q)^2}
 \exp \left \{-  \frac{(\bb p_{Ti}-\bb q)^2}{2 \sigma_u^2}  \right \}
\nonumber\\
&&=
\frac{1}{(\sqrt{2\pi} \sigma)^2}
\exp \left \{- \frac{\bb p_{Ti}^2}{2 \sigma^2}  \right \} ,
\label{eq17}
\end{eqnarray}
where 
\begin{eqnarray}
\sigma^2=\sigma_u^2+\sigma_q^2.
\end{eqnarray}
Therefore, when the four constituents $p_1,p_2,p_3,p_4$ are produced
independently without correlations for two nonadjacent mesons, the
transverse momentum distribution $dN_T^N$ of these particles is
\begin{eqnarray}
{dN_T^N}\!\!=\!\!
 \frac
{d\bb p_{T1} d\bb p_{T2}d \bb p_{T3} d\bb p_{T4} }{(\sqrt{2 \pi} \sigma)^8 }   
\exp  \!\left \{ \!
- \frac{\bb p_{T1}^2\! +\! \bb p_{T2}^2 \!+\!\bb p_{T3}^2\! +\! \bb p_{T4}^2}{2\sigma^2 }\!\right  \}\!\!.
\nonumber\\
\label{19}
\end{eqnarray}
  For the most likely case in the production of mesons with light
  quarks with $m_1$=$m_2$=$m_3$=$m_4$, the transformations (\ref{3})
  and (\ref{12b}) from $(\bb p_{Ti}$,$\bb p_{Tj})$ to $(\bb
  P_{Tij}$,$\bb p_{T ij})$ can be easily carried out.  In terms of the
  composite transverse momenta and their relative momenta, the
  momentum distribution (\ref{19}) becomes
\begin{eqnarray}
&&\hspace*{-0.7cm}{dN_T^N}=
 \frac{1}{(\sqrt{2 \pi} \sigma)^8 }   
{d \bb P_{T12}  d \bb P_{T34} d \bb p_{T12} d \bb p_{T34}}
\nonumber\\
&& \hspace*{-0.5cm}\times \exp\left \{ \!\! - \frac{1}{2\sigma^2} \left ( \frac{\bb P_{T12}^2}{2} + 2 \bb p_{T12}^2
+ \frac{\bb P_{T34}^2}{2} + 2 \bb p_{T34}^2 \right ) \!\!\right \}\!.
\end{eqnarray}
The relative transverse momenta $\bb p_{T12}$ and $\bb p_{T34}$ can be
integrated out to yield
\begin{eqnarray}
{dN_T^N}=&&
 \frac{{d \bb P_{T12}  d \bb P_{T34} }}{4(\sqrt{2 \pi} \sigma)^4 }   
\exp\left \{ \!\! - \frac{1}{2\sigma^2} \left ( \frac{\bb P_{T12}^2}{2}\!  + \!\frac{\bb P_{T34}^2}{2} 
 \right ) \!\!\right \}.
\end{eqnarray}
  We represent $\bb P_{Tij}$ by $(p_{Tij},\phi_{ij})$ and introduce  
the azimuthal angle difference $\Delta \phi$ and sum $\Sigma$, 
\begin{eqnarray}
\Delta \phi&& = \phi_{12}-\phi_{34},~~~~ \Sigma =  \phi_{12}+\phi_{34}.
\end{eqnarray}
Upon averaging over $\Sigma$, we have
\begin{eqnarray}
&&dN_T^N=
\frac{1}{4(\sqrt{2 \pi} \sigma)^4 } \frac{1}{2} d\Delta \phi  \frac{1}{\Sigma_{\rm max}-\Sigma_{\rm min}} \int_{\Sigma_{\rm min}}^{\Sigma_{\rm max}} d\Sigma
\nonumber\\
&&\times\!\!
 \int \!\!d P_{T12} d P_{T12}   P_{T34} d P_{T34} 
\exp\left \{ - \frac{1}{2\sigma^2} \left (\frac{P_{T12}^2 +P_{T34}^2}{2} \right ) \!\right  \}\!,
\nonumber
\end{eqnarray}
which gives
\begin{eqnarray}
\frac{dN_T^N}{d\Delta \phi}=\frac{1}{8\pi^2}.
\label{eq23}
\end{eqnarray}
Consequently, for
nonadjacent mesons in the absence of any correlation between the
produced quarks and antiquarks, the transverse momentum distribution
$dN_T^N$ is independent of the azimuthal angle difference $\Delta
\phi$ of two nonadjacent mesons.

\section{The Two-Particle Transverse  Distribution 
for Adjacent Mesons}

For two adjacent mesons
as shown in
Fig.\ \ref{fig2}, the transverse distribution $dN_T^N$ in Eq.\ (\ref{19}) in the last
section will need to be modified to become
\begin{eqnarray}
&&dN_T^A=dN_q dN_{q'} dN_{q''}
 \frac{1}{(\sqrt{2 \pi} \sigma_u)^{8} }
  d\bb p_{T1} d\bb p_{T2} d\bb p_{T3} d\bb p_{T_4}
\nonumber\\
&&\times \!\exp \!\{\! - \frac{(\bb p_{T1}\!\!-\!\bb q')^2\! +\! (\bb p_{T2}\!\!-\!\bb q)^2 \!+\!(\bb p_{T3}\!+\!\bb q) ^2\! +\! (\bb p_{T4}\!-\!\bb q'')^2}{2\sigma_u^2} \},
\nonumber\\
\label{22n}
\end{eqnarray}
where the
momentum kicks $\bb q$, $\bb q'$, and $\bb q''$  of the Schwinger mechanism 
 have the distributions
\begin{eqnarray}
dN_{q_i}\!\! = \!\! \frac{1}{(\sqrt{2\pi}\sigma_q)^2}
d\bb q_i \, \exp\{ - \frac{\bb q_i^2}{2\sigma_q^{2} }\},~~{\bb q_i}\!=\!\! \bb q, \bb q', {\rm and~}\bb  q'' \!\!.~~~~~
\label{20a}
\end{eqnarray}
The quark
$p_2$ and the antiquark $p_3$ 
in Fig.\ \ref{fig2}
are produced by the
Schwinger mechanism at the point $V$.  Because of the local
conservation of momentum at the production point $V$, the transverse 
momenta
$\bb p_{T2}$ and $\bb p_{T3}$ are correlated and the correlation is expressed in Eq.\ (\ref{22n})
as a shift of their momenta  $\bb p_{T2}$ and $\bb p_{T3}$ in Eq.\ (\ref{19}) to $ \bb p_{T2} -\bb q$ and $ \bb p_{T3} +\bb q$ respectively. 
 The constituents $p_1$ and $p_4$ themselves
are products of Schwinger pair production and their transverse momenta $ \bb p_{T1}$  and $ \bb p_{T4}$
become $ \bb p_{T1} -\bb q'$, and $ \bb p_{T4} -\bb q''$, 
depending on the momentum kicks $\bb q'$ and $\bb q''$.

After integrating over $\bb q'$ and $\bb q''$ in the distributions $dN_{q'}$ and $dN_{q''}$ as in Eq.\ (\ref{eq17}),  the momentum distribution
$dN_T^A$ in Eq.\ (\ref{22n}) becomes
\begin{subequations}
\begin{eqnarray}
&&dN_T^A=dN_q d{\tilde N}_T(q),
\\
&&d{\tilde N}_T(q)= \frac{1}{(\sqrt{2 \pi} \sigma)^{4} }\frac{1}{(\sqrt{2 \pi} \sigma_u)^{4} }
  d\bb p_{T1} d\bb p_{T2} d\bb p_{T3} d\bb p_{T4}
\nonumber\\
&&\hspace*{0.3cm}\times \exp\left \{ - \frac{\bb p_{T1}^2}{2\sigma^2}
 +\frac{ (\bb p_{T2}-\bb q)^2 +(\bb p_{T3}+\bb q) ^2 }{2\sigma_u^2}
 + \frac{\bb p_{T4}^2}{2\sigma^2}
\right  \}.~~~~~~~
\label{24b}
\end{eqnarray}
\end{subequations}
Our task is to separate out the relevant degrees of freedom in
Eq.\ (\ref{24b}).  The most likely case is the production of mesons
with light quarks, for which we can take $m_1$=$m_2$=$m_3$=$m_4$.  By
using the momentum transformations of Eqs. (\ref{3}) and ({\ref{12b}),
  we convert $\bb p_{T1}$, $\bb p_{T2}$, $\bb p_{T3}$, and $\bb p_{T4}$, to $\bb p_{T12}$, $\bb p_{T34}$,
$\bb P_{T12}$, and  $\bb P_{T34}$.  Equation  (\ref{24b}) becomes
\begin{eqnarray}
&&d{\tilde N}_T(q)= \frac{1}{(\sqrt{2 \pi} \sigma)^{4} }\frac{1}{(\sqrt{2 \pi} \sigma_u)^{4} }
  d\bb P_{T12} d\bb P_{T34} d\bb p_{T12} d\bb p_{T34}
\nonumber\\
&&\hspace*{0.3cm}\times \exp\left \{ 
A+B+C+D+E
\right  \}.~~~~~~~
\label{25}
\end{eqnarray}
where we collect  the quadratic function of $\bb p_{T12}$ in $A$,
\begin{eqnarray}
A\!=\!-\!\bb p_{T12}^2 [ \frac{1}{2\sigma^2}\!+\!\frac{1}{2\sigma_u^2}] 
\!-\! \bb p_{T12}\!\cdot\! [\frac{\bb P_{T12}}{2\sigma^2}\!+\!\frac{-\bb P_{T12} + 2\bb q}{2\sigma_u^2}],~~~~
\label{eq26}
\end{eqnarray}
we collect  the quadratic function of $\bb p_{T34}$ in $B$,
\begin{eqnarray}
B\!=\!-\bb p_{T34}^2 [ \frac{1}{2\sigma^2}\!+\!\frac{1}{2\sigma_u^2}] 
\!-\! \bb p_{T34}\!\cdot\! [\frac{\!-\!\bb P_{T34}}{2\sigma^2}\!+\!\frac{+\bb P_{T34} \!+\! 2\bb q}{2\sigma_u^2}],~~~~
\label{eq27}
\end{eqnarray}
we collect  the remaining quadratic function of $\bb P_{T12}$ in $C$,
\begin{eqnarray}
C=-\bb P_{T12}^2 \frac{1}{4}\cdot[  \frac{1}{2\sigma^2}+\frac{1}{2\sigma_u^2}] 
- \bb P_{T12}\cdot [\frac{-\bb q}{2\sigma_u^2}],
\end{eqnarray}
we collect  the remaining quadratic function of $\bb P_{T34}$ in $D$,
\begin{eqnarray}
D=-\bb P_{T34}^2 \frac{1}{4}\cdot[  \frac{1}{2\sigma^2}+\frac{1}{2\sigma_u^2}] 
- \bb P_{T34}\cdot [\frac{+\bb q}{2\sigma_u^2}],
\end{eqnarray}
and the last term quadratic in $\bb q^2$ in $E$, 
\begin{eqnarray}
E=- \frac{2\bb q ^2 }{2\sigma_u^2}.
\end{eqnarray}
By completing the squares of $A$ and $B$ in Eqs.\ (\ref{eq26}) and (\ref{eq27}), the integration over $\bb p_{T12}$ and $\bb p_{T34}$  can be carried out analytically, yielding a function of $\bb P_{T12}, \bb P_{T34}$, and $\bb q$.  After these integrations, we obtain 
\begin{eqnarray}
&&d{\tilde N}(q) 
=\frac{1}{(2\pi)^2(\sigma^2+\sigma_u^2)^2}
   d\bb P_{T12} d\bb P_{T34} 
\nonumber\\
&&\times 
\exp \biggl \{ \frac{\sigma_{12}^2}{2}\left [ \frac{-\sigma_q^2 \bb P_{T12}}{2 \sigma^2 \sigma_u^2}
\!+\! \frac{\bb q}{\sigma_u^2} \right ]^2
\!\!\!+\! \frac{\sigma_{12}^2}{2}\left  [ 
 \frac{+\sigma_q^2 \bb P_{T34}}{2 \sigma^2 \sigma_u^2}  
\!+\! \frac{\bb q}{\sigma_u^2}  \right ]^2
\biggr  \}
\nonumber\\
&&\times
\exp \left \{ \!
- \frac{\bb P_{T12}^2}{8\sigma_{12}^2} 
\!+\!  \frac{\bb P_{T12}\cdot\bb q}{2\sigma_u^2}
\!-\! \frac{\bb P_{T34}^2}{8\sigma_{12}^2}
\!-\!  \frac{\bb P_{T34}\cdot\bb q}{2\sigma_u^2}
\!-\! \frac{\bb q ^2 }{\sigma_u^2}
\!\right \},
\end{eqnarray}
where
\begin{eqnarray}
\frac{1}{2\sigma_{12}^2}= \frac{1}{2\sigma^2}+\frac{1}{2\sigma_u^2} . \nonumber
\end{eqnarray}•
For the remaining degrees of freedom, we define $\phi_{12}$ as the
azimuthal angle between $\bb q$ and $\bb P_{T12}$, and $\phi_{34}$ as
the azimuthal angle between $\bb q$ and $\bb P_{T34}$.  We form the
sum and the difference of the azimuthal angles
\begin{subequations}
\begin{eqnarray}
\Delta \phi&& = \phi_{12}-\phi_{34},~~~~ \Sigma =  \phi_{12}+\phi_{34},\\
\phi_{12}&&=\frac{1}{2}(\Sigma+ \Delta \phi  ), ~~~~ \phi_{34}=\frac{1}{2}( \Sigma-\Delta \phi  ). \label{21bb}
\end{eqnarray}
\end{subequations}
We then get the distribution $d {\tilde N}_T(q)$ defined in the
intervals $-2\pi$$\le$$ \Delta\phi \le 2\pi$ and $\Sigma_{\rm
  min}(\Delta \phi)$$ \le $$\Sigma \le \Sigma_{\rm max}(\Delta \phi)$
where
\begin{subequations}
\begin{eqnarray}
\Sigma_{\rm min}(\Delta \phi) &&=  - 2\pi +|\Delta \phi|,
\\
\Sigma_{\rm max}(\Delta \phi) &&=-|\Delta \phi|+2\pi.
\end{eqnarray}
\end{subequations}
\begin{figure}[h]
 \centering
\includegraphics[scale=0.45]{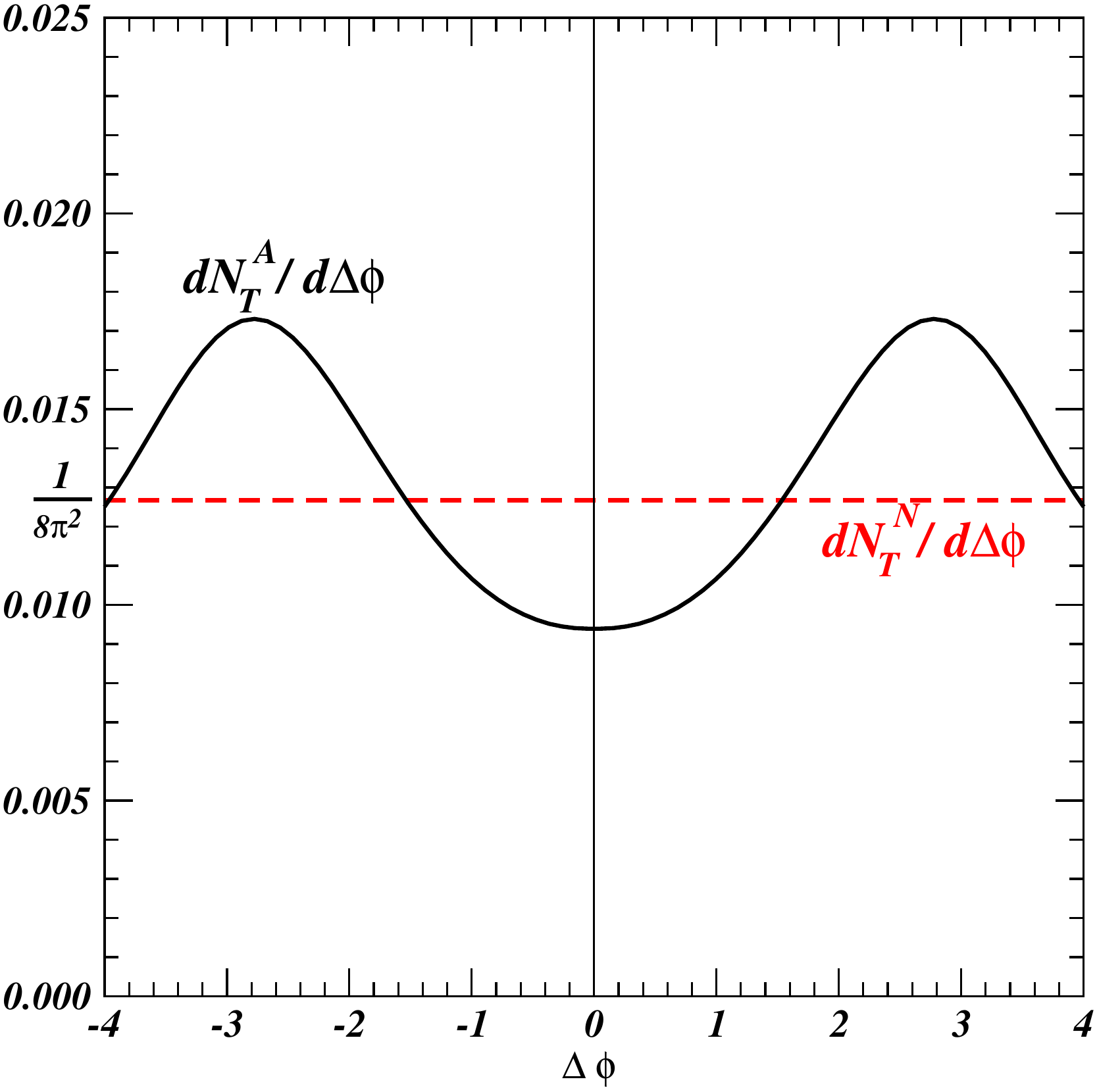}
\caption{ (Color online) The transverse momentum distributions
   as a function of the
  difference $\Delta \phi$ of the azimuthal angles between two
  adjacent mesons,    $dN_T^A/d\Delta \phi$ , and 
two nonadjacent mesons,
 $d N_T^N/d\Delta \phi$ .
}
\label{corr}
\end{figure}
The normalized distribution per pair of mesons $P_{12}$ and $P_{34}$ is
\begin{eqnarray}
&&\frac{d{\tilde N}_T(q)}{d\Delta \phi}(\!\Delta \phi\!)
 =\! \frac{1}{(\Sigma_{\rm max}\!\!\!-\!\Sigma_{\rm min})}
\!\!\int_{\Sigma_{\rm min}} ^ {\Sigma_{\rm max}} \!\!\!\!\!\!\!\!\!d\Sigma
\frac{ P_{T12} dP_{T12}
P_{T34} dP_{T34}
} {2(2 \pi)^2( \sigma^2+\sigma_u^2)^2}
\nonumber\\
&&\times 
\exp \biggl \{ \frac{\sigma_{12}^2}{2}\left [ \frac{-\sigma_q^2 \bb P_{T12}}{2 \sigma^2 \sigma_u^2}
\!+\! \frac{\bb q}{\sigma_u^2} \right ]^2
\!\!\!+\! \frac{\sigma_{12}^2}{2}\left  [ 
 \frac{+\sigma_q^2 \bb P_{T34}}{2 \sigma^2 \sigma_u^2}  
\!+\! \frac{\bb q}{\sigma_u^2}  \right ]^2
\biggr  \}
\nonumber\\
&&\times
\exp \left \{ \!
- \!\frac{\bb P_{T12}^2}{8\sigma_{12}^2} 
\!+\!  \frac{\bb P_{T12}\cdot\bb q}{2\sigma_u^2}
\!-\! \frac{\bb P_{T34}^2}{8\sigma_{12}^2}
\!-\!  \frac{\bb P_{T34}\cdot\bb q}{2\sigma_u^2}
\!-\! \frac{\bb q ^2 }{\sigma_u^2}
\right \}.
\label{eq34}
\end{eqnarray}
where the scalar product  $\bb P_{Tij} \cdot \bb q$
is
\begin{eqnarray}
\bb P_{Tij} \cdot \bb q=P_{Tij} ~q ~\cos(\phi_{ij}).
\end{eqnarray} 
Here,
$\phi_{12}$ and $\phi_{34}$ in the integrand can be expressed as
a function of $\Delta \phi$ and $\Sigma$ as given by
Eq.\ (\ref{21bb}).  We can calculate $d{\tilde N}_T(q)/d\Delta \phi$
by integrating out $\Sigma, P_{T12}$, and $P_{T34}$ in the above
equation, and we obtain $d{\tilde N}_T(q)/d\Delta \phi$ for different
values of $q$. 
We need the values of 
$\sigma_u$, and $\sigma_q$ to get the numerical estimate.  For a flux-tube radius
of $R_{\rm tube}=0.6$ fm and $\kappa=1$ GeV/fm, we get
$\sigma_u=0.33$ GeV/c, $\sigma_q=0.25$ GeV/c, and $\sigma$=0.414 GeV/c. 

One finds that the distribution $d{\tilde N}_T(q)/d\Delta \phi$
depends on the magnitude of the momentum $|\bb q|/\sigma_q$ whose
distribution is given by Eq.\ (\ref{20a}).  
Upon integrating the
distribution of $q$, we obtain the distribution of $\Delta \phi$ for
two adjacently produced mesons $P_{12}$ and $P_{34}$, 
\begin{eqnarray}
\frac{dN_T^A}{d\Delta \phi}&&=\int  \frac{d\bb q}{(\sqrt{2\pi} \sigma_q)^2 } \exp\{ - \frac{\bb q^2}{2\sigma_q^2} \}  \frac{d{\tilde N}_T(q)}{d\Delta \phi}.
\label{24a}
\end{eqnarray}
The solid curve in Fig.\ \ref{corr} is the $d{ N}_T^A/d \Delta \phi$
result obtained by a direct numerical integration over $q$.  The
correlation is suppressed on the near side at $\Delta \phi$$\sim$0,
and is enhanced on the back-to-back away side at $\Delta
\phi$$\sim$$\pi$.  Such a result is expected because the quark $p_2$
and the antiquark $p_3$ are produced in opposite azimuthal directions
due to the local momentum conservation.  The associated adjacently
produced mesons $P_{12}$ and $P_{34}$ should also have back-to-back
azimuthal correlations.

Along with the correlation for an adjacent pair of mesons, there is
also the azimuthal angular correlation $dN_T^{N}/d\Delta \phi$ for a
pair of nonadjacent mesons in a flux-tube fragmentation.  The angular
correlation $dN_T^N/d\Delta \phi$ corresponds to the case where the
quark $p_2$ and the antiquark $p_3$ are not correlated by the momentum
$\bb q$.  
It is given by Eq.\ (\ref{eq23})
with $dN_T^N/d\Delta \phi$=1$/8\pi^2$.  It can also be alternatively obtained by setting $q$=0 in Eq.\ (\ref{eq34}). 

The above results have been obtained by applying the local law of
conservation of momentum for a system of equal quark masses.  As the
constituent quark mass of a strange quark is only slightly greater
than the constituent quark masses of the up and down quarks \cite{Won01}, the
qualitative feature of the above results regarding a back-to-back
correlation for adjacent meson production may also be applicable to
cases involving the production of strange mesons.  

\section{ Rapidity Correlation of Produced Mesons in a Flux-Tube Fragmentation}

The last section gave the azimuthal correlation of two adjacent and
nonadjacent mesons in a flux-tube fragmentation.  How do we identify
adjacent and nonadjacent pairs of mesons?  We shall use the
space-time-rapidity ordering to correlate the rapidities of produced
mesons with the spatial locations on the flux tube at the moment of
fragmentation.
 
Casher $et~al.$ \cite{Cas74} showed that in QED2 when a quark parton
pulls away from an antiquark with the speed of light, the produced
dipole density of produced $q$-$\bar q$ pairs is a Lorentz-invariant
function and the lines of constant produced dipole density of produced
$q$-$\bar q$ pairs are hyperbolas with constant proper times.  It is
therefore reasonable to assume as a first approximation that the
space-time distribution of the $q$-$\bar q$ production vertices should
depend only on the vertex production proper time, $\tau_{\rm pro}$,
relative to the point of the onset of separation of the leading quark
and antiquark.

Previously in Ref.\ \cite{Won91b} and in Exercise (7.1) of Ref.\  \cite{Won94}, we
showed that if all the pair-production vertices of a fragmenting string
fall on the curve of the proper time $\tau_{\rm pro}$, the rapidity
distribution of the produced mesons is a
constant given by
\begin{eqnarray}
\frac{dN}{dy}=\frac{\kappa \tau_{\rm pro}}{m_T},
\label{26}
\end{eqnarray}
where $\kappa$ is the string tension and $m_T$ is the transverse mass
of the particle $m_T$=$\sqrt{m^2+p_T^2}$ where $m$ is the mass of a
produced meson (mostly likely a pion).  Thus the presence of a
rapidity plateau is an indication of the approximate validity of the
occurrence of the the $q$-$\bar q$ vertices along a curve of constant
proper time.  We showed further in Exercise (7.2) of Ref.\  \cite{Won94} that
in that case, the produced particles are ordered in
space-time-rapidity.  The ordering of produced particles means that in
the center-of-mass system, particles with a greater magnitude of
rapidity $|y|$ are produced at a distance farther away from the point
of collision and at a time later than and more separated from the time
of collision, as shown in Fig.\ \ref{order}.
\begin{figure}[h]
 \centering
\includegraphics[scale=0.50]{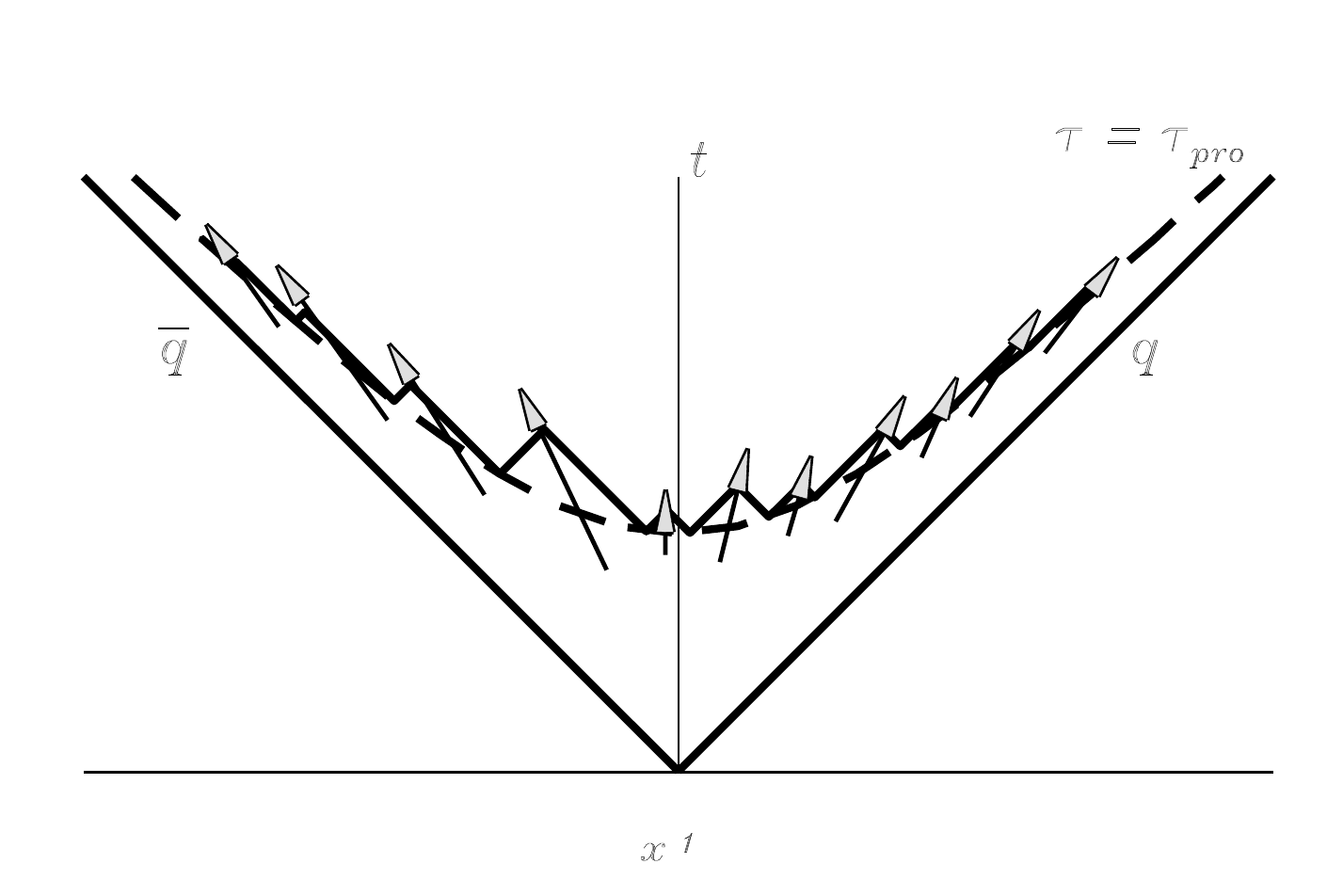}
\caption{ (Figure taken from Figure 6.4 of \cite{Won94}).  The
  space-time trajectories of the leading quark $q$ moving to the
  positive $x^1$ direction (longitudinal direction) and the antiquark
  moving to the negative $x^1$ direction, with the production of
  $q$-$\bar q$ pairs lying on the proper time curve of constant
  $\tau_{pro}$.  A produced $q$ at one vertex interacts with the
  produced $\bar q$ in the adjacent vertex to form a meson.  The
  trajectories of mesons are indicated by thick arrows.  The
  rapidities of the produced mesons are ordered along the spatial
  longitudinal $x^1$ axis,  and in time.  }
\label{order}
\end{figure}
From such a picture of flux-tube fragmentation, we envisage that
produced mesons are ordered in space-time and in rapidity, with mesons
that are produced adjacent to each other closest in their magnitudes
of their rapidity values \cite{Won94}.  Therefore, mesons are
produced adjacent to each other if their rapidity difference $\Delta
y$ falls within a rapidity width $ w$,
\begin{eqnarray}
|\Delta y |\lesssim w,
\end{eqnarray}
where $w$ is the rapidity per produced meson, 
\begin{eqnarray}
w = \frac{1}{dN/{dy} } \sim \frac{1}{ (3/2)dN_{ch}/ dy}.
\label{eq42}
\end{eqnarray}
One notes from Eq.\ (\ref{26}) that the width $w$ is inversely
proportional to the particle production time $\tau_{\rm pro}$.  As the
particle production time cannot be sharp and has a distribution (see Fig. 2 of \cite{Won91b} for an example of the scatter plot of the pair-production vertex proper times in the Lund model), the
width $w$ would also have a distribution.  The condition relating
$\Delta y$ with adjacently produced mesons should also have a
diffuseness described by an additional parameter $a$.  Therefore,
within the nearest-neighbor rapidity width $|\Delta y|$$\lesssim$$w$ in the
central rapidity region, the produced mesons are adjacent to each
other  and the rapidity correlation for an adjacent pair of mesons is
\begin{eqnarray}
\frac{dN_y^A}{d\Delta y}= \frac{1}{1+e^{(|\Delta y|-w)/a}}.
\end{eqnarray}

Upon approximating the rapidity $y$ as the pseudorapidity $\eta$, the
above correlation for adjacent pair becomes
\begin{eqnarray}
\frac{dN_\eta^A}{d\Delta \eta}= \frac{1}{1+e^{(|\Delta \eta|-w)/a}}.
\label{eq43}
\label{37}
\end{eqnarray}
The complete correlation consists of both the transverse and
longitudinal correlations.  From the transverse azimuthal correlation
$dN_T^A/d\Delta \phi$ for adjacent mesons as given in Fig.\ \ref{corr}, we have therefore
the two-particle angular correlation for an adjacent pair of mesons as
\begin{eqnarray}
\frac{dN^{A}}{d\Delta \phi~ d\Delta \eta} =\frac{dN_T^{A}}{d\Delta \phi}~\frac{dN_\eta^{A}}{d \Delta \eta}
=\frac{1}{1+e^{(|\Delta \eta|-w)/a}}\frac{dN_T^{A}}{d\Delta \phi}.\hspace*{0.5cm}
\label{A}
\end{eqnarray}

On the other hand, outside this nearest-neighbor rapidity width with
$|\Delta \eta|$$\lesssim$$w$, the two mesons are nonadjacent.  The
condition of being a nonadjacent pair, outside the nearest-neighbor
rapidity width with $|\Delta \eta|$$\lesssim$$ w$ can be specified by
$1/{(1+e^{(w-|\Delta \eta|)/a})}$.  Consequently, the two-particle
pseudorapidity correlation for a nonadjacent pair of mesons in a flux
tube fragmentation is
\begin{eqnarray}
\frac{dN_\eta^N}{d \Delta \eta} =&& \frac{1}{1+e^{(w-|\Delta \eta|)/a}}.
\end{eqnarray}
From the transverse azimuthal correlation $dN_T^N/d\Delta \phi$ as
given in Eq.\ (\ref{eq23}), we have therefore the two-particle angular
correlation for a nonadjacent pair of mesons as
\begin{eqnarray}
\frac{dN^N}{d\Delta \phi~ d\Delta \eta}&& =\frac{dN_T^N}{d\Delta \phi}~\frac{dN_\eta^N}{d \Delta \eta}
=\frac{1}{1+e^{(w-|\Delta \eta|)/a}}\frac{1}{8\pi^2}. \hspace*{0.5cm}
\label{N}
\end{eqnarray}

In experimental measurements, one selects a pair of mesons whose
charges and heavy-quark quantum numbers can be the same or opposite.
Different selections will result in different adjacent and nonadjacent
fractions, and consequently different linear combinations of $dN^{A}/
{d\Delta \phi \, d\Delta \eta}$ and $dN^N/ {d\Delta \phi \, d\Delta
  \eta}$.  They will lead to different shapes of the correlation
functions.  It is necessary to find out the adjacent and nonadjacent 
fractions for different charge and heavy-quark configurations.

\section{Charge Correlation in Flux-Tube Fragmentation for Quarks with Two Flavors} 
\vspace*{0.3cm}

In a flux-tube fragmentation, the production of the quark-antiquark
pair $p_2$ and $p_3$ at the local point $V$ in Fig.\ \ref{fig2} must
satisfy local charge and flavor conservations.  This means that $p_2$
and $p_3$ must be quark and antiquark with opposite charges
and flavor quantum numbers.  Such a local conservation will
induce correlations on charges and flavors of adjacent 
mesons.  For nonadjacent mesons, there will be no such  correlations.

We can investigate all possible charge and flavor contents of two
produced mesons if they are in the adjacent or nonadjacent state.  We
study the case of quarks with two flavors in this section, and the
case with three flavors in the next section.

We examine first the production of two adjacent mesons and construct
in Table \ref{tb1} all possible meson production configurations in
which the quark $p_2$ and the antiquark $p_3$ have different
combinations of charges and flavor quantum numbers.  Here, $Q_{ij}$ is
the charge of the composite meson $P_{ij}$ with constituents $p_i$ and
$p_j$.
 
There are altogether eight possible cases in Table \ref{tb1}.  The
number of cases $N^A(Q_{12},Q_{34})$ and the associated probabilities
$P^A(Q_{12},Q_{34})$ in different charge configurations
$(Q_{12},Q_{34})$ for two adjacent mesons in a flux-tube fragmentation
are given in Table \ref{t2}.  We note that because $p_2$ and $p_3$
have equal and opposite charges, two adjacently produced mesons
$P_{12}$ and $P_{34}$ cannot have charges of the same sign as shown in
Table \ref{t2}.

\begin{table}[H]
\centering
\caption { Quark and antiquark configurations for two adjacently
  produced mesons in a flux-tube fragmentation where $p_2$ and $p_3$
  are constrained by local charge conservation and flavor balance.
 }
\label{tb1}
\vspace*{0.2cm}
\begin{tabular}{|c|c|c|c|c|c|}
\cline{1-6}
     $p_1$    &  $p_2$  &  $Q_{12}$  & $p_3$  &  $ p_4$  &  $Q_{34}$  
  \\
\cline{1-6}
$\bar u$  & $u$ &  0 & $\bar u$  & $d$ & -1 
\\
$\bar d$  &$u$ &  1 & $\bar u$  & $d$ & -1 
\\ 
$\bar u$  & $u$ &  0 & $\bar u$  & $u$ & 0 
\\ 
$\bar d$  &$u$ &  1 & $\bar u$  & $u$ & 0 
\\
$\bar u$  & $d$ &  -1 & $\bar d$  & $d$ & 0 
\\ 
$\bar d$  & $d$ & 0 & $\bar d$  & $d$ & 0 
\\
$\bar u$  & $d$ &  -1 & $\bar d$  & $u$ & 1 
\\ 
$\bar d$  &$d$ &  0 & $\bar d$  & $u$ & 1 
\\ \hline
\end{tabular}
\end{table}

\begin{table}[H]
\centering
\caption { The number of cases $N^A(Q_{12},Q_{34})$ and the probability $P^A(Q_{12},Q_{34})$ of the charge configurations for two adjacently
  produced mesons in a flux-tube fragmentation. }
\label{t2}
\vspace*{0.2cm}
\begin{tabular}{|c|c|c|c|c|}
\cline{3-5}
  \multicolumn{1}{c}       {}  &   &  $Q_{34}$=-1  &  $Q_{34}=0$  &  $Q_{34}$=1 \\ 
\cline{1-5}
 &$Q_{12}$=-1  &    0    &   1  &  1 
\\
 $N^A(Q_{12},Q_{34})$ & $Q_{12}$=~0  & 1    &   2    & 1
\\ 
  & $Q_{12}$=+1  &    1 &  1     & 0
\\ 
\cline{1-5}
 &$Q_{12}$=-1  &    0    &   0.125  &  0.125 
\\
 $P^A(Q_{12},Q_{34})$ & $Q_{12}$=~0  & 0.125    &   0.250    & 0.125
\\ 
  & $Q_{12}$=+1  &    0.125 &  0.125      & 0
\\ \hline
\end{tabular}
\end{table}

We next examine the production of two mesons in the nonadjacent
state.  We construct in Table \ref{t3} all charge configurations of
two nonadjacent mesons.  In that case, the charge and flavor quantum
numbers of quark $p_2$ and antiquark $p_3$ need not be correlated.
Without the constraint on the flavors of  $p_2$ and $p_3$, one has 
altogether 16 possible cases as listed in
Table \ref{t3}.  The number of cases $N^N(Q_{12},Q_{34})$ and the
associated probabilities $P^N(Q_{12},Q_{34})$ in different charge
configurations $(Q_{12},Q_{34})$ for two nonadjacent mesons in a flux
tube fragmentation are given in Table \ref{t4}.  In this case, 
there is an equal probability for two nonadjacent mesons
$P_{12}$ and $P_{34}$ to have charges of the same sign or opposite
signs.  This is in contrast to adjacent mesons for which there is no
probability that their charges will be of the same sign.

\begin{table}[H]
\centering
\caption { Quark and antiquark configurations for two nonadjacent 
mesons  in flux
  tube fragmentation where $p_2$ and $p_3$ are not constrained by 
  flavor balance and local charge conservation. }
\label{t3}
\vspace*{0.2cm}
\begin{tabular}{|c|c|c|c|c|c|}
\cline{1-6}
     $p_1$    &  $p_2$  &  $Q_{12}$  & $p_3$  &  $ p_4$  &  $Q_{34}$  
  \\
\cline{1-6}
$\bar u$  & $u$ &  0 & $\bar u$  & $d$ & -1 
\\
$\bar d$  &$u$ &  1 & $\bar u$  & $d$ & -1 
\\ 
$\bar u$  & $u$ &  0 & $\bar u$  & $u$ & 0 
\\ 
$\bar d$  &$u$ &  1 & $\bar u$  & $u$ & 0 
\\
$\bar u$  & $d$ &  -1 & $\bar d$  & $d$ & 0 
\\ 
$\bar d$  & $d$ & 0 & $\bar d$  & $d$ & 0 
\\
$\bar u$  & $d$ &  -1 & $\bar d$  & $u$ & 1 
\\ 
\vspace*{0.1cm}
$\bar d$  &$d$ &  0 & $\bar d$  & $u$ & 1 
\\
$\bar u$  & $u$ &  0 & $\bar d$  & $d$ & 0 
\\
$\bar d$  &$u$ &  1 & $\bar d$  & $d$ &0 
\\ 
$\bar u$  & $u$ &  0 & $\bar d$  & $u$ & 1
\\ 
$\bar d$  &$u$ &  1 & $\bar d$  & $u$ &1 
\\
$\bar u$  & $d$ &  -1 & $\bar u$  & $d$ & -1
\\ 
$\bar d$  & $d$ & 0 & $\bar u$  & $d$ & -1 
\\
$\bar u$  & $d$ &  -1 & $\bar u$  & $u$ & 0
\\ 
$\bar d$  &$d$ &  0 & $\bar u$  & $u$ & 0
\\ \hline
\end{tabular}
\end{table}

\begin{table}[H]
\centering
\caption { The number of cases $N^N(Q_{12},Q_{34})$ and the
  probability $P^N(Q_{12},Q_{34})$ of all charge configurations for
  two nonadjacent mesons $P_{12}$ and $P_{34}$ in a flux-tube
  fragmentation. }
\label{t4}
\vspace*{0.2cm}
\begin{tabular}{|c|c|c|c|c|}
\cline{3-5}
  \multicolumn{1}{c}       {}  &  & $Q_{34}=-1$ &    $Q_{34}=0$  &  $Q_{34}$=1 \\ 
\cline{1-5}
 &$Q_{12}$=-1  &    1   &   2  &  1 
\\
 $N^N(Q_{12},Q_{34})$ & $Q_{12}$=~0  & 2    &   4    & 2
\\ 
  & $Q_{12}$=+1  &    1 &  2     & 1
\\ 
\cline{1-5}
&$Q_{12}$=-1  &    0.0625    &   0.125 &   0.0625 
\\
$P^N(Q_{12},Q_{34})$&$Q_{12}$=~0  &  0.125     &     0.250    & 0.125
\\ 
&$Q_{12}$=+1  &    0.0625  &    0.125     & 0.0625
\\ \hline
\end{tabular}
\end{table}
The charge pattern for two produced mesons in the adjacent state is
quite different from the charge pattern of mesons in the nonadjacent
state, in a flux-tube fragmentation.

\section{Charge and Flavor Correlation in Flux-Tube Fragmentation for 
Quarks with Three Flavors} 

\vspace*{0.3cm}

The considerations in the last section for the production of $u$-$\bar
u$ and $d$-$\bar d$ pairs can be generalized to the case with the
additional production of $s$-$\bar s$ pairs. Compared to the
production of a $u$ or $d$ quark pair, the probability for the
production of a strange-quark pair is suppressed by a strangeness
suppression factor $f_s$ that is energy dependent
\cite{Mak15,Gaz15,Lar15,Ser15,Adakpi04,Wol11,PHO11,Vov14}, and is of
order 0.10 at $ \sqrt{s_{_{NN}}}$= 200 GeV as shown in Appendix A.
We can enumerate all charge configurations in a flux-tube
fragmentation, for both adjacent and nonadjacent mesons.  The
additional strangeness degree of freedom brings in an additional
correlation of the strange quantum numbers of the produced mesons.

In enumerating all possible configurations for the production of 
two mesons, it is useful to associate each configuration by the order
of the strangeness suppression factor $f_s$.  We use the light  $q$-$\bar q$ 
pair-production probability as  a unit of measure of order 1.  Upon 
referring to the constituents in Fig.\ \ref{fig2} for adjacent mesons
and Fig.\ \ref{fig3} for nonadjacent mesons, the production of an
$\bar s$ in $p_1$ or an $s$ quark in $p_4$ will each bring in a
suppression factor of order $f_s$.  Similarly, the production of an
$s$ in $p_2$ and an $\bar s$ in $p_3$ in two nonadjacent mesons will
also each bring in a factor of $f_s$.  However, the production of an
$s$ in $p_2$ and an $\bar s$ in $p_3$ in adjacent mesons will bring in
altogether only a single factor of $f_s$ because this $s$-$\bar s$
pair arises from a single Schwinger pair-production mechanism.  From
these considerations, we can construct all possible meson charges
$(Q_{12},Q_{34})$ and meson strangeness $(S_{12},S_{34})$ up to first
order in $f_s$ in Table V for adjacent meson pairs.  Here, $S_{ij}$ is
the strangeness of the composite meson $P_{ij}$ with constituents
$p_i$ and $p_j$.

\begin{table}[H]
\centering
\label{t5}
\caption { Quark, antiquark, charge, and strangeness configurations
  for two adjacent mesons in a flux-tube fragmentation, where
  $p_2$ and $p_3$ are constrained by charge and flavor balance,
  for flux-tube fragmentation with three flavors.  }
\vspace*{0.2cm}
\begin{tabular}{|c|c|c|c|c|c|c|c|c|}
\cline{1-9}
     $p_1$    &  $p_2$  &  $Q_{12}$  &  $S_{12}$  & $p_3$  &  $ p_4$  &  $Q_{34}$ &  $S_{34}$ 
& order
  \\
\cline{1-9}
$\bar u$  & $u$ &  0 &  0 & $\bar u$  & $d$ & -1 &  0 & 1 
\\
$\bar d$  &$u$ &  1 &  0 & $\bar u$  & $d$ & -1 &  0 & 1 
\\ 
$\bar u$  & $u$ &  0 &  0 & $\bar u$  & $u$ & 0 &  0 & 1 
\\ 
$\bar d$  &$u$ &  1 &  0 & $\bar u$  & $u$ & 0 &  0 & 1 
\\ 
$\bar u$  & $d$ &  -1 &  0 & $\bar d$  & $d$ & 0 &  0 & 1 
\\ 
$\bar d$  & $d$ & 0 &  0 & $\bar d$  & $d$ & 0 &  0 & 1 
\\
$\bar u$  & $d$ &  -1 &  0 & $\bar d$  & $u$ & 1 &  0 & 1 
\\ 
$\bar d$  &$d$ &  0 &  0 & $\bar d$  & $u$ & 1 &  0 & 1 
\\
\hline
$\bar u$  & $s$ &  -1 &  -1 &$\bar s$  & $d$ & 0 & 1& $f_s$
\\ 
$\bar d$  & $s$ & 0 & -1& $\bar s$  & $d$ & 0 & 1& $f_s$
\\
$\bar u$  & $s$ &  -1 & -1& $\bar s$  & $u$ & 1 & 1& $f_s$
\\ 
$\bar d$  &$s$ &  0 & -1& $\bar s$  & $u$ & 1 & 1& $f_s$
\\ 
$\bar s$  &$u$ &  1 & 1 & $\bar u$  & $d$ & -1 &  0 & $f_s$

\\
$\bar s$  &$u$ &  1 & 1 & $\bar u$  & $u$ & 0 &  0 & $f_s$
\\
$\bar u$  & $u$ &  0 &  0 & $\bar u$  & $s$ & -1 & -1& $f_s$
\\
$\bar d$  &$u$ &  1 &  0 & $\bar u$  & $s$ & -1 & -1& $f_s$
\\
$\bar s$  & $d$ & 0 & 1 & $\bar d$  & $d$ & 0 &  0 & $f_s$

\\ 
$\bar s$  &$d$ &  0 & 1 & $\bar d$  & $u$ & 1&  0 & $f_s$
\\ 
$\bar u$  & $d$ &  -1 &  0 & $\bar d$  & $s$ & 0 & -1& $f_s$
\\ 
$\bar d$  & $d$ & 0 &  0 & $\bar d$  & $s$ & 0 & -1& $f_s$
\\\hline
\end{tabular}
\end{table}
 
From the above Table V, we can construct the number of different
charge configurations in Table \ref{t6} for two adjacent mesons in 
  a flux-tube fragmentation with three
  flavors.   Here,  the superscript $(ud)$
specifies the $ud$ sector, and $(s)$ the strange sector.

\begin{table}[H]
\caption { Up to order $f_s$, the number of cases
  $N^{A(ud)}(Q_{12},Q_{34})$ and $N^{A(s)}(Q_{12},Q_{34})$ for
  different charge configurations $(Q_{12},Q_{34})$, in two adjacent
  mesons $P_{12}$ and $P_{34}$ in a flux-tube fragmentation with three
  flavors.
}
\label{t6}
\vspace*{0.2cm}
\begin{tabular}{|c|c|c|c|c|}
\cline{3-5}
  \multicolumn{1}{c}       {}    &  &  $Q_{34}$=-1  &  $Q_{34}=0$  &  $Q_{34}$=1 \\ 
\cline{1-5}
 &$Q_{12}$=-1  &    0    &   1  &   1  
\\
$N^{A(ud)}(Q_{12},Q_{34})$ & $Q_{12}$=~0  &  1      &     2   & 1
\\ 
  & $Q_{12}$=~1  &    1  &     1    &  0
\\ \hline
\cline{1-5}
 &$Q_{12}$=-1  &    0    &   2$ f_s$  &    $ f_s$
\\
 $N^{A(s)}(Q_{12},Q_{34})$ & $Q_{12}$=~0 &  $f_s$     &     3$ f_s$   &  2$ f_s$
\\ 
 & $Q_{12}$=~1  &   2$ f_s$  &   $ f_s$     & 0
\\ \hline
\end{tabular}
\end{table}

In a flux-tube fragmentation, there is a local conservation of
strangeness or heavy-quark flavor in the production of a $Q$-$\bar Q$
pair in the flux tube, similar to the conservation of charge.  Hence,
there are also correlations that arise from strangeness and heavy-quark 
conservation in a flux-tube fragmentation.  We present in Table
\ref{t7} the number of cases for the production of strangeness
configurations $(S_{12},S_{34})$, in a flux-tube fragmentation with three
  flavors.

\begin{table}[H]
\centering
\caption { The number of cases $N^{A(ud)}(S_{12},S_{34})$ and
  $N^{(ud)}(S_{12},S_{34})$ for different strangeness configurations
  $(S_{12},S_{34})$, for two adjacent mesons $P_{12}$ and $P_{34}$ in
  a flux-tube fragmentation with three flavors.  }
\label{t7}
\vspace*{0.2cm}
\begin{tabular}{|c|c|c|c|c|}
\cline{3-5}
  \multicolumn{1}{c}       {}    &  &  $S_{34}$=-1  &  $S_{34}=0$  &  $S_{34}$=1 \\ 
\cline{1-5}
&$S_{12}$=-1  &    0    &   0  &    $4 f_s$
\\
$N^{A(s)}(S_{12},S_{34})$ &  $S_{34}$=~0  &  $4f_s$     &     0  &  0
\\ 
  & $S_{12}$=~1  &   0  &   4$ f_s$     & 0
\\ \hline
\end{tabular}
\end{table}

From the above Table \ref{t6}, the probability
$P^{A(ud,s)}(Q_{12},Q_{34})$ for the occurrence of adjacent
non-strange mesons $(ud)$ and strange mesons $(s)$, with charges
$(Q_{12},Q_{34})$ is related to the number of cases
$N^{A(ud,s)}(Q_{12},Q_{34})$ in flux-tube fragmentation with three flavors
by
\begin{eqnarray}
P^{A(ud,s)}(Q_{12},Q_{34}) = \frac{N^{A(ud,s)}(Q_{12},Q_{34})}{8+12 f_s},
\label{43}
\end{eqnarray}
which is normalized to
\begin{eqnarray}
\sum_{Q_{12},Q_{34}}[P^{A(ud)}(Q_{12},Q_{34})+P^{A(s)}(Q_{12},Q_{34})] =1.~~~~
\end{eqnarray}
Similarly, if one considers the strangeness configurations of two
adjacent mesons, the probability $P^{A(ud,s)}(S_{12},S_{34})$ for the
occurrence of strange mesons with strangeness $(S_{12},S_{34})$ is
related to the number of cases $N^{A(ud,s)}(S_{12},S_{34})$ in flux-tube fragmentation with three flavors
by

\begin{eqnarray}
P^{A(s)}(S_{12},S_{34}) = \frac{N^{A(s)}(S_{12},S_{34})}{8+12 f_s}.
\label{45}
\end{eqnarray}

In a similar manner, we can investigate all possible quark,
antiquark, charge, and strangeness configurations for two nonadjacent
mesons $P_{12}$ and $P_{34}$ in a flux-tube fragmentation.  In that
case, the charge and flavor of $p_3$ will not need to be correlated
with those of $p_2$.  The details of the enumeration is presented in
Appendix B.  Again, we need to keep track of the order of the
strangeness suppression factor $f_s$ in different configurations.
Keeping configurations only up to the first order in $f_s$, we list
as a summary the number of cases $N^{N(ud)}$ and $N^{N(s)}$ for different
combinations of charge and strangeness configurations in Tables
\ref{tbl8} and \ref{tbl9} respectively.
 
\begin{table}[H]
\centering
\caption { Up to order $f_s$, the number of cases
  $N^{N(ud)}(Q_{12},Q_{34})$ and $N^{N(s)}(Q_{12},Q_{34})$ of the
  charge configurations for two nonadjacent mesons $P_{12}$ and
  $P_{34}$ in a flux-tube fragmentation with three flavors.
}
\vspace*{0.3cm}
\label{t8}
\begin{tabular}{|c|c|c|c|c|}
\cline{3-5}
  \multicolumn{1}{c}       {}    &  &  $Q_{34}$=-1  &  $Q_{34}=0$  &  $Q_{34}$=1 \\ 
\cline{1-5}
 &$Q_{12}$=-1  &    1   &   2   &   1 
\\
$N^{N(ud)}(Q_{12},Q_{34})$  & $Q_{12}$=~0  &  2      &     4   & 2
\\ 
  & $Q_{12}$=~1  &    1   &     2    &  1
\\ \hline
\cline{1-5}
 &$Q_{12}$=-1  &    2$f_s$    &   4$f_s$   &     2$f_s$
\\
$N^{N(s)}(Q_{12},Q_{34})$ & $Q_{12}$=~0  &  4$f_s$    &     8$f_s$   &  4$f_s$
\\ 
  & $Q_{12}$=~1  &   2$f_s$  &    4$f_s$     & 2$f_s$
\\ \hline
\end{tabular}
\label{tbl8}
\end{table}

\begin{table}[H]
\centering
\caption { The number of cases $N^{N(ud)}(S_{12},S_{34})$ and
  $N^{N(ud)}(S_{12},S_{34})$ of the strangeness configurations of two
  nonadjacent mesons in a flux-tube fragmentation with three flavors.
}
\label{t9}
\begin{tabular}{|c|c|c|c|c|}
\cline{3-5}
  \multicolumn{1}{c}       {}    &  &  $S_{34}$=-1  &  $S_{34}=0$  &  $S_{34}$=1 \\ 
\cline{1-5}
 &$S_{12}$=-1  &    0    &   8$f_s$   &     0
\\
$N^{N(s)}(S_{12},S_{34})$ & $S_{12}$=~0  &  8$f_s$     &     0  &  8$f_s$
\\ 
   & $S_{12}$=~1  &   0  &    8$f_s$    &  0
\\ \hline
\end{tabular}
\label{tbl9}
\end{table}

From the above Table \ref{t8}, we can sum up all possible numbers of cases, and
get the probability $P^{N(ud,s)}(Q_{12},Q_{34})$ for nonadjacent
mesons.  The probabilities for the occurrence of nonstrange mesons
$(ud)$ and strange mesons $(s)$, with charges $(Q_{12},Q_{34})$
for a flux-tube fragmentation with three flavors 
 is
related to the number of cases $N^{N(ud,s)}(Q_{12},Q_{34})$ by
\begin{eqnarray}
P^{N(ud,s)}(Q_{12},Q_{34}) = \frac{N^{N(ud,s)}(Q_{12},Q_{34})}{16+32 f_s},
\label{46}
\end{eqnarray}
which is normalized to
\begin{eqnarray}
\sum_{Q_{12},Q_{34}}[P^{N(ud)}(Q_{12},Q_{34})+P^{N(s)}(Q_{12},Q_{34})] =1.~~~~
\end{eqnarray}

We also have the probability $P^{N(ud,s)}(S_{12},S_{34})$ for
nonadjacent mesons with strangeness $ (S_{12},S_{34})$ 
for flux-tube fragmentation with three flavors 
given by
\begin{eqnarray}
P^{N(ud,s)}(S_{12},S_{34}) = \frac{N^{N(ud,s)}(S_{12},S_{34})}{16+32 f_s}.
\label{48}
\end{eqnarray}

\section{Two-Meson Correlation Function in Different Charge  
and Strangeness Configurations }

The results in the last few sections allow us to provide different
signatures for a color flux-tube fragmentation.  One measures the
$\Delta \phi$-$\Delta \eta$ correlations for a given charge or
strangeness configuration $\nu$ within a $p_T$ domain of interest, the
pattern of the correlation will reveal the nature of the reaction
mechanism within that $p_T$ domain, as in the case for the
hard-scattering process for $p_T$$>$0.5 GeV/c in Fig. 1.

To specify a charge or strangeness configuration $\nu$, one can choose
a pair of mesons with electric charges of the same sign, for which
$Q_{12}$=$Q_{34}$, of opposite sign, for which $Q_{12}$=$-Q_{34}$,
with the same strangeness, for which $S_{12}$=$S_{34}$, with opposite
strangeness, for which $S_{12}$=$-S_{34}$, or the correlation of a
strange meson with a nonstrange meson. Different choices will yield
different probabilities $P^A$ and $P^N$ for finding adjacent or
nonadjacent pair of mesons, and will lead to different patterns of
the correlation function, in the fragmentation of a color flux tube.

We can write down the two-meson distribution function $dN/ d\Delta
\phi d\Delta \eta$ for a pair of mesons with azimuthal angle
difference $\Delta \phi$ and pseudorapidity difference $\Delta \eta$
in a charge configuration (or strangeness configuration) $\nu$ within
a given $p_T$ domain in flux-tube fragmentation as given by
\begin{eqnarray}
\frac{dN(\nu)}{d\Delta \phi d\Delta \eta} =&&
\frac{dN^A}{d\Delta \phi d\Delta \eta}   \times  P^A(\nu)
+  \frac{dN^N}{d\Delta \phi d\Delta \eta} \times P^N(\nu) . 
\nonumber\\
\label{49}
\end{eqnarray}
Here, the first factor in each term, ${dN^X}/{d\Delta \phi d\Delta
  \eta}$, is the distribution  function as a function of  $\Delta \phi$ and $\Delta \eta$,
when the meson pair is
an adjacent pair with $X$=$A$ or a nonadjacent pair with $X$=$N$, in
the fragmentation of the flux tube, as given by Eqs.\ (\ref{A}) and (\ref{N}).  The second factor $P^X(\nu)$ in
each term specifies the probability for the pair of mesons in the $X$
state to be in the configuration $\nu$ of interest.  The total
probability $P^X(\nu)$ for the charge configuration $\nu$ in meson
pairs in the state $X$ is given by
\begin{eqnarray}
P^X(\nu)=\!\!\!\! \sum_{(Q_{12},Q_{34})\in  \nu } \!\!\!\! [P^{X(ud)}(Q_{12},Q_{34}) +  P^{X(s)}(Q_{12},Q_{34})] . 
\nonumber
\\
\end{eqnarray}
The total probability $P^X(\nu)$ for the strangeness configuration
$\nu$ to be in the state $X$ is given by
\begin{eqnarray}
P^X(\nu)=\sum_{(S_{12},S_{34})\in  \nu }  [P^{X(ud)}(S_{12},S_{34}) +  P^{X(s)}(S_{12},S_{34})].  
\nonumber\\
\end{eqnarray}
The quantities $P^{X(ud,s)}(S_{12},S_{34})$ and $P^{X(ud,s)}(Q_{12},Q_{34})$ 
are given by Eqs.\ (\ref{43})-(\ref{48}) and Tables
\ref{t6}-\ref{t9}.

We can consider the configuration $\nu$ of two mesons with charges of
opposite signs.  We obtain from Table \ref{t6} for adjacent mesons and Eq.\ (\ref{43}),
\begin{eqnarray}
P^A(\text{opposite charge}) \!=\! \frac{2+3f_s}{8+12 f_s}\!=\!\frac{1}{4},~~~~~~
\end{eqnarray}
and from Table \ref{t8} for nonadjacent mesons and Eq.\ (\ref{46}),
\begin{eqnarray}
P^N(\text{opposite charge}) \!=\! \frac{2+4f_s}{16+32 f_s}\!=\!\frac{1}{8}.~~~~~~
\end{eqnarray}
Similarly, for the charge configuration $\nu$ of two mesons with
charges of the same sign, we obtain from Table \ref{t6} for adjacent
mesons and Eq.\ (\ref{43})
\begin{eqnarray}
P^A(\text{same charge}) =0,
\end{eqnarray}
and from Table \ref{t8}  for nonadjacent mesons and Eq.\ (\ref{46}), 
\begin{eqnarray}
P^N(\text{same charge}) =\frac{2+4f_s}{16+32f_s}=\frac{1}{8}.
\end{eqnarray}
These probabilities for different charge configurations are
independent of $f_s$, indicating that they can be obtained by
considering either two flavors or three flavors.  On the other hand,
the probability for different strangeness configurations must be
obtained by considering three flavors.

We list below the probabilities $(P^A(\nu), P^N(\nu))$ for a few
two-meson configurations $\nu$:
\begin{enumerate}
\item
$\nu$=(two mesons with charges of opposite signs),
\begin{eqnarray}
(P^A,P^N)(\text{opposite charge}) = (0.25,~0.125).
\label{54}
\end{eqnarray}
\item
$\nu$=(two mesons with charges of the same sign),
\begin{eqnarray}
(P^A,P^N)(\text{same charge}) = (0,~0.125).~~~~~~
\label{55}
\end{eqnarray}

\item
$\nu$=(two charged mesons ($Q_{12}\ne 0$ and $Q_{34}\ne 0$)),
\begin{eqnarray}
(P^A,P^N)(\text{all charge}) = (0.25,~0.25),~~~~~~
\label{56}
\end{eqnarray}
which is the sum of the first  two cases.

\item
$\nu$=(two mesons with opposite strangeness),
\begin{eqnarray}
(P^A,P^N)(\text{opposite strangeness}) = (\frac{4f_s}{8+12f_s},0).~~
\label{57}
\end{eqnarray}

\item
$\nu$=(two mesons with same strangeness),
\begin{eqnarray}
(P^A,P^N)(\text{same strangeness}) = (0,0),
\end{eqnarray}
 which is true only to order $f_s$.  The probabilities will be non-vanishing
to the next order of  $f_s^2$.

\item
$\nu$=(one $|S|$=1 meson and one nonstrange  meson),
\begin{eqnarray}
&&(P^A,P^N)(\text{one strange, one nonstrange})
\nonumber\\
&& \hspace*{1.5cm}= (\frac{8 f_s}{8+12f_s}, 
\frac{32 f_s}{16+32f_s} ).
\label{59}
\end{eqnarray}

\end{enumerate}

\section{Numerical Examples of Theoretical $dN/d\Delta \phi d\Delta \eta$ 
in Flux-Tube Fragmentation}

Using Eqs.\ (\ref{A}), (\ref{N}), and (\ref{49}), we obtain the
correlation function for two mesons in a charge or strangeness
configuration $\nu$ as given by
 \begin{eqnarray}
\frac{dN(\nu)}{d\Delta \phi  d\Delta \eta}&& =\! P^A\!(\nu) \frac{1}{1+e^{(|\Delta \eta|-w)/a}}
\frac{dN_T^{A}}{d\Delta \phi}
\nonumber\\
&& 
+ P^N(\nu) \frac{1}{1+e^{(w - |\Delta \eta|)/a}} \frac{1}{8\pi^2},
\label{60}
\end{eqnarray}
where $dN_T^{A}/d\Delta \phi$ is given by Eq.\ (\ref{24a}) and its numerical values are shown in Fig.\ \ref{corr}.

In typical experimental measurements of the  two-hadron angular correlations
such as those in the NA61/SHINE Collaboration
\cite{Mak15,Gaz15,Lar15,Ser15}, one collects 
the data for a set of events with similar characteristics (such as charge multiplicities) and picks a pair of mesons of  configuration $\nu$   (for example, of opposite charges) from  the same event to build up the ``sibling" distribution  ${dN_{\rm sib}(\nu)}/{d\Delta \phi  d\Delta \eta}$ per meson pair, as a function of $\Delta \phi$ and $\Delta \eta$.  To eliminate systematic errors and to account for the phase space boundaries of the detectors, one also uses this set of data to pick
a pair of   mesons of   configuration $\nu$
 from two different events (mixed events)  to 
  build up the ``mixed" distribution ${dN_{\rm mix}(\nu)}/{d\Delta \phi  d\Delta \eta}$ per meson pair  as a function of $\Delta \phi$ and $\Delta \eta$.   The sibling distribution contains all the correlation information while the meson pair from different ``mixed" events do not contain any correlation.  We can construct the sibling-to-mixed ratio 
\begin{eqnarray}
\frac{{dN_{\rm sib}(\nu)}/{d\Delta \phi  d\Delta \eta}}
{{dN_{\rm mix}(\nu)}/{d\Delta \phi  d\Delta \eta}},
\label{eq68}
\end{eqnarray}
which can be compared with theoretical  predictions $C(\Delta \phi,\Delta \eta)$ for
such a ratio. 

The numerator of (\ref{eq68}) can be identified with the distribution of Eq.\ (\ref{60}).   The denominator from different events in an event mixing,
can be considered in a hypothetical case if the meson pair of configuration $\nu$ arises in the absence of 
all  correlations. 
With the absence of correlations
in $\phi$, then  from  Eq.\ (\ref{eq23}) $dN_\phi/d\Delta \phi$=$1/8\pi^2$,
and with  the absence of space-time-ordering correlations in $\eta$, then  
$dN_\eta/d\Delta \eta$=$dN_\eta^A/d\Delta \eta $+$dN_\eta^N/d\Delta \eta$=1.   The distribution ${dN_{\rm mix}(\nu)}/{d\Delta \phi  d\Delta \eta}$ in the absence of all correlations is $[P^A\!(\nu)$+$P^N\!(\nu) ] /{8\pi^2}$.   It is therefore useful to 
divide the distribution (\ref{60}) by the scale $[P^A\!(\nu)$+$P^N\!(\nu) ] /{8\pi^2}$
to construct the following theoretical angular correlation function for comparison with experimental ratio of Eq.\ (\ref{eq68}) 
 \begin{eqnarray}
C^{\rm single ~flux~tube}(\Delta \phi, \Delta \eta)
 =\frac{\frac{dN(\nu)}{d\Delta \phi  d\Delta \eta}}
 {[P^A\!(\nu) + P^N\!(\nu) ]  \frac{1}{8\pi^2}},~~~~~
\label{68}
\end{eqnarray}
where $dN(\nu)/d\Delta \phi  d\Delta \eta$ is given by Eq.\ (\ref{60}).

It is however necessary to make  an amendment  in the above equation in order to compare with the flux-tube fragmnentaion in $pp$ collisions.   The above result gives theoretical  correlation function for a single flux tube.  In flux-tube fragmentation in $pp$ collisions, two flux tubes are formed by the diquark of one nucleon forming a flux tube with the valence quark of the other nucleon, and vice versa.  The mesons from one flux-tube fragmentation will not be correlated with the mesons from the other flux-tube fragmentation, but they are included in the sibling and mixed distributions in the experimental measurement.   These uncorrelated mesons from two different flux tubes
contribute $P^N(\nu) /8\pi^2$ to both the sibling and the mixed distributions. Upon taking into account this contribution from two different flux tubes, the proper theoretical angular correlation function for the charge or strangeness
configuration $\nu$ in $pp$ collisions is 
 \begin{eqnarray}
C(\Delta \phi, \Delta \eta)
\! =\!\frac{
\frac{P^A\!(\nu) }{1+e^{(|\Delta \eta|-w)/a}}
\frac{dN_T^{A}}{d\Delta \phi}
\!+\! \frac{P^N(\nu) }{1+e^{(w - |\Delta \eta|)/a}} \frac{1}{8\pi^2}
\!+\! \frac{P^N(\nu) }{8\pi^2}}
 {[P^A\!(\nu) + P^N\!(\nu) ]  \frac{1}{8\pi^2}  +P^N\!(\nu)   \frac{1}{8\pi^2}  },
\nonumber
\\
\label{eq69}
\end{eqnarray}
which is to be compared with the experimental sibling-to-mixed ratio of Eq.\ (\ref{eq68}).

We can consider an explicit numerical example.  For $pp$ collisions at
$\sqrt{s_{_{NN}}}$=200 GeV, we have $dN_{\rm ch}/d\eta|_{\eta\sim0}
\sim$ 2.25 \cite{Adakpi04,Wol11,PHO11}.  As $pp$ collisions contains
contributions from two quark-diquark strings, and each string produces
$dN_{ch}/d\eta$$ \sim$1.125, we therefore have $w $$\sim$0.59 at
$\sqrt{s}=200$ GeV.

We can estimate the diffuseness parameter $a$ from the ratio of the
width (represented by the standard deviation $\sigma_\tau$) of the
distributions of the pair production proper time to the average pair
production proper time $\langle \tau_{\rm pro} \rangle$.  Previously
in the scatter plot of the pair production proper times in the Lund
model (see Fig. 2 of \cite{Won91b}), we find
\begin{eqnarray}
\frac{\sigma_\tau}{\langle \tau_{\rm pro} \rangle} \sim \frac {1}{2}.
\end{eqnarray}
From Eqs.\ (\ref{26}) and (\ref{eq42}),  $w$ is inversely proportional to the pair production proper time $\tau_{\rm pro}$.  Hence, we have
\begin{eqnarray}
\frac{\text{(width of }w)}{w} = \frac{ \text{(width of }\tau_{\rm pro})}{\langle \tau_{\rm pro}\rangle}.
\end{eqnarray}
The width of $w$  can be identified 
with 1.5$a$, for which the Fermi distribution of Eq.\ (\ref{eq43}) drops from  82\% to 0.18\% at $w$$-$$1.5a$ to $w$+$1.5a$.  We obtain then 
\begin{eqnarray}
\frac{1.5 a}{w} \sim \frac{\sigma_{\tau}}{\langle \tau_{\rm pro} \rangle}\sim \frac{1}{2}, ~~~\text{and}~~ a\sim 0.2.
\end{eqnarray}
For numerical purposes, we shall use $w=0.59$ and $a$=$0.2$ in subsequent
calculations.

\subsection{ Mesons with Opposite Charges}

For the case of a pair of mesons with opposite charges, the
probabilities for the pair to be in the adjacent and nonadjacent
states, as given by Eq.\ (\ref{54}), are $(P^A,P^N)$=(0.25,0.125).

\begin{figure}[H]
 \centering
\includegraphics[scale=0.75]{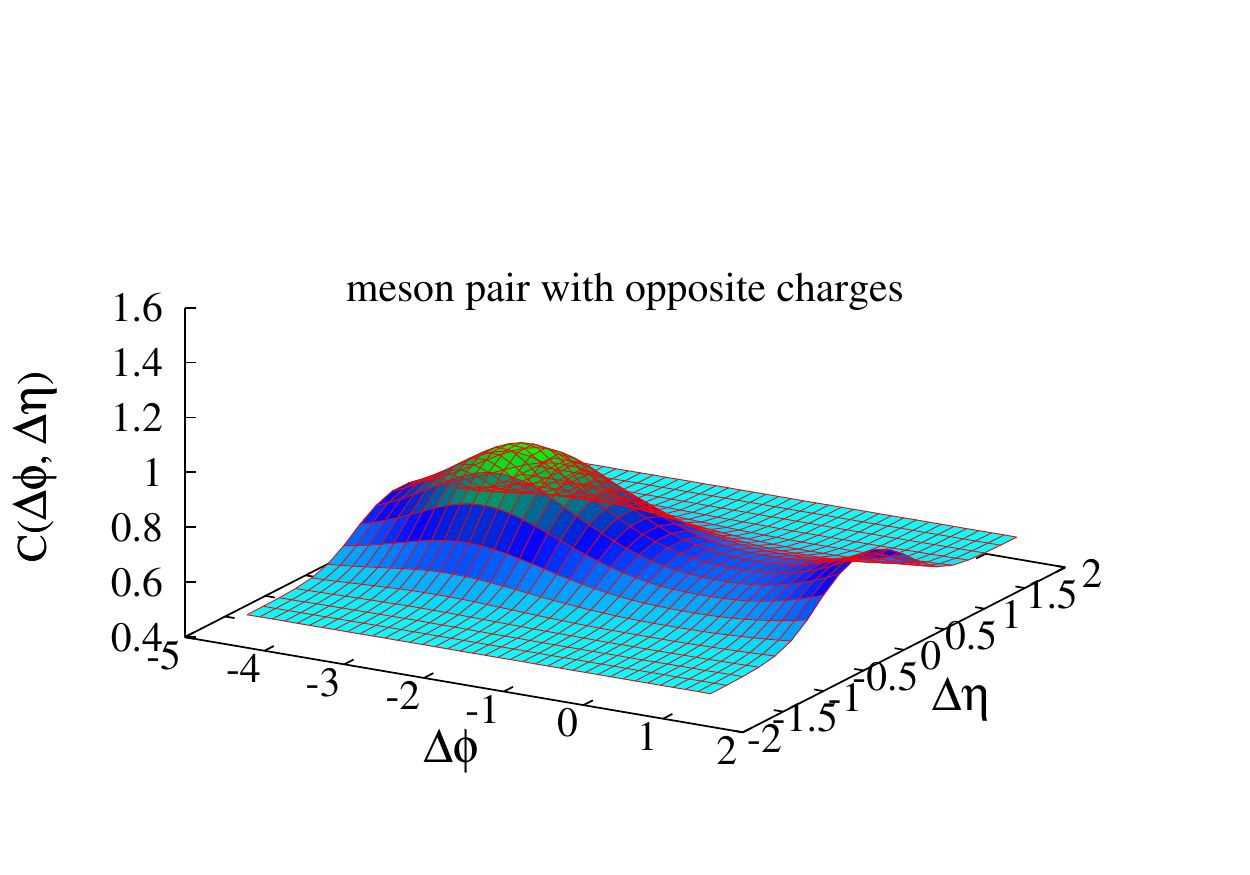}
\caption{ (Color online) The two-meson angular correlation $C(\Delta \phi,\Delta \eta)$
 for two oppositely charged mesons  in a flux-tube fragmentation
in $pp$ collisions.  }
\label{opp}
\end{figure}

In
Fig.\ \ref{opp}, the contributions in Eq.\ (\ref{eq69}) combine
together to yield the two-meson angular correlation $C(\Delta \phi,\Delta \eta)$, 
in a flux-tube fragmentation for two mesons with
opposite charges, calculated for the sample case of $w$=0.59 and
$a$=0.2 in $pp$ collisions.  
We observe that the correlation is suppressed at $(\Delta
\phi, \Delta \eta)\sim 0$ but is significantly enhanced at $(\Delta
\phi$$\sim $$\pi, \Delta \eta$$ \sim$0). 
 Such a correlation arises
from the conservation of momentum, charge, and flavor in the
production of a quark and an antiquark that become constituents of the
two adjacent mesons.  For two adjacent mesons that arise from the
nearest neighbors, their momenta are likely to be azimuthally
correlated to be back-to-back because a constituent of one of the
mesons and a constituent of the other meson share the same 
pair-production process at the same spatial point that requires local
conservation of momentum.  Their charges are likely to have opposite
signs because the constituent of one of the mesons and the constituent
of the other meson share the same pair-production process that
requires local charge conservation.  

\subsection{ Mesons with Charges of the Same Sign}

For the measurement on a pair of mesons with charges of the same sign,
the probabilities for the pair to be in the adjacent and nonadjacent
states, as given by Eq. (\ref{55}), are $(P^A,P^N)$=(0,0.125).
Because $P^A=0$, the correlation is independent of $\Delta \phi$. 

\begin{figure}[H]
 \centering
\includegraphics[scale=0.75]{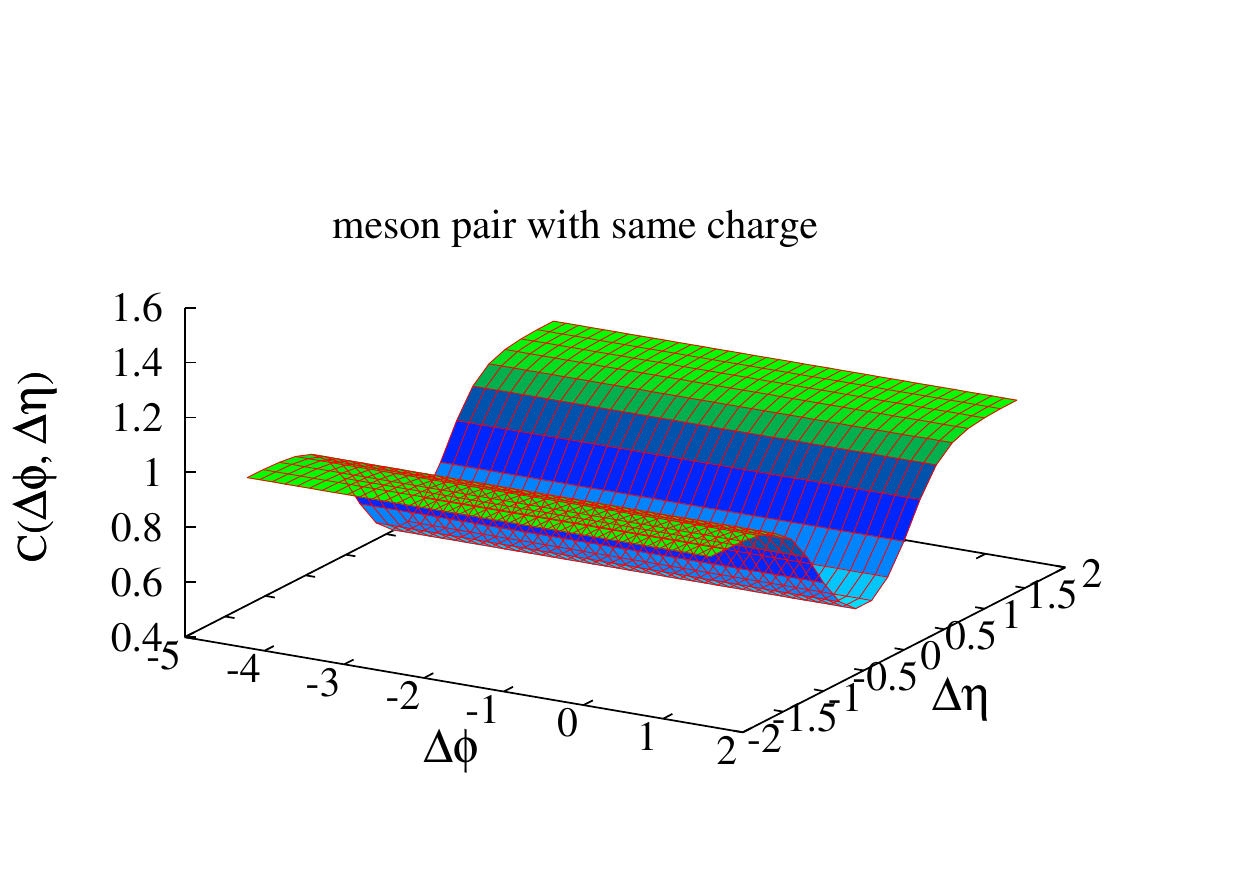}
\caption{ (Color online) The two-meson angular correlation 
$C(\Delta \phi,\Delta \eta)$  for meson pairs
  with charges of the same sign  in a flux-tube fragmentation in $pp$ collisions.    }
\label{same}
\end{figure}

 We
show in Fig.\ \ref{same} the correlation function 
 $C(\Delta \phi,\Delta \eta)$
  in a flux-tube fragmentation for a pair of mesons with
charges of the same sign.  We observe a suppression at $\Delta \eta \sim 0$
indicating that few, if any, pairs of mesons can be produced with
charges of the same sign at $\Delta \eta\sim0$.  Such a suppression
arises because $\Delta \eta \sim 0$ signals the meson to be adjacently
produced but mesons produced adjacently are forbidden to have charges
of the same sign, as indicated in Table \ref{t6}.

\subsection{Charged  Meson  Pair}

In another example of charge combinations, one detects a pair of
charged mesons and measures the correlation between one charged meson
with another charged meson.  Then from Eq.\ (\ref{56}), we have
$(P^A,P^N)$=(0.25,0.25).  There is an equal probability for the two
mesons to be an adjacent pair or a nonadjacent pair.

\begin{figure}[H]
 \centering
\includegraphics[scale=0.75]{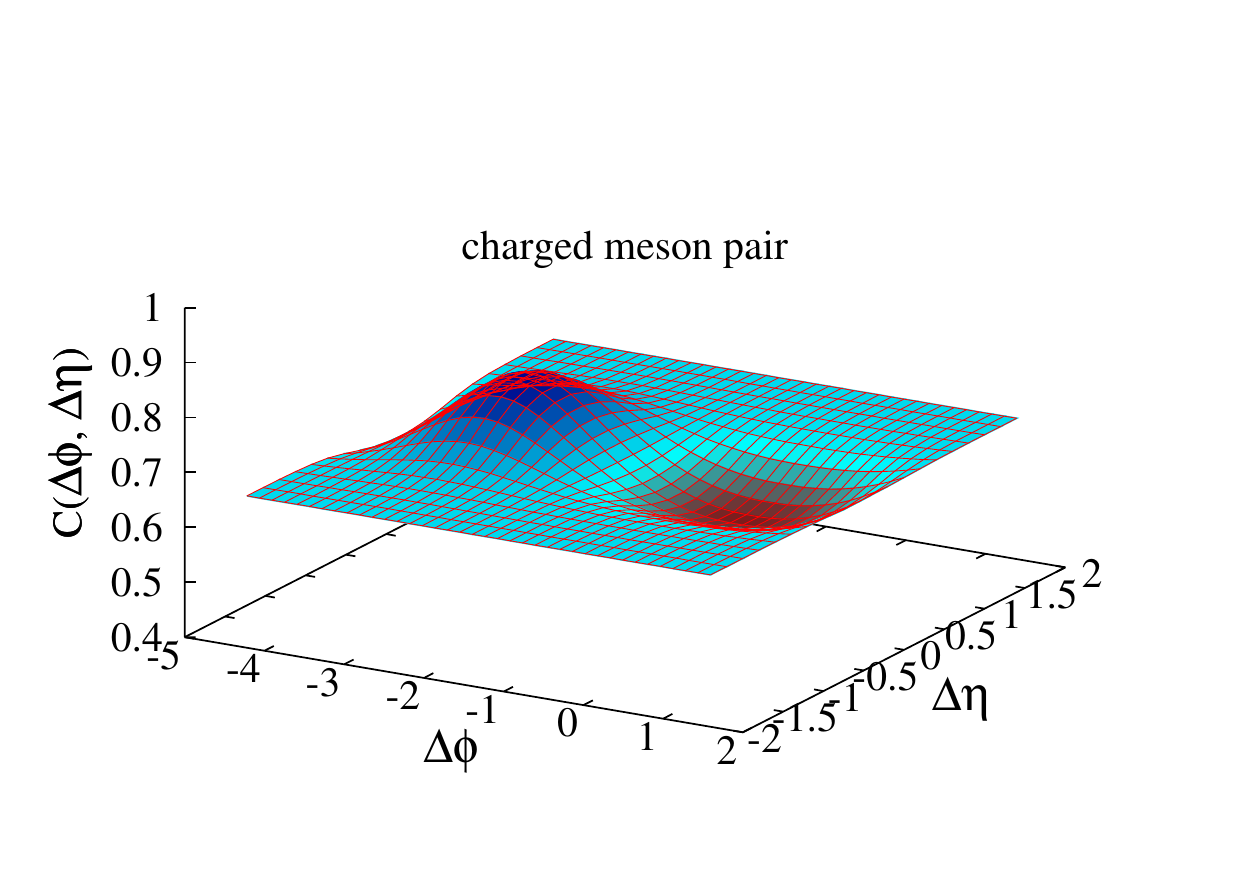}
\caption{ (Color online) The two-meson angular correlation
 $C(\Delta \phi,\Delta \eta)$  for
  charged meson pairs in a flux-tube fragmentation in $pp$ collisions.
}
\label{all}
\end{figure}

We show the two-meson angular correlation $C(\Delta \phi,\Delta \eta)$
 for two charged mesons for the case of flux-tube fragmentation
in Fig.\ \ref{all}.  It is in fact related  to the sum of the correlation
functions of Figs.\ \ref{opp} and \ref{same}.  The suppression at
$(\Delta \phi,\Delta \eta)$$ \sim$0 arises from the suppression of
charges both of the same sign and opposite signs at $(\Delta
\phi,\Delta \eta)$$ \sim$0, while the enhancement at $(\Delta
\phi$$\sim$$\pi,\Delta \eta$$\sim $0) arises from the back-to-back enhanced 
correlation of charges of opposite signs in adjacent mesons.

\subsection{Mesons with Opposite Strangeness}

For a pair of mesons with opposite strangeness, Eq.\ (\ref{57}) gives
the probability $P^A=f_s/(2+3f_s)$ for the pair to be in the adjacent
state.  The probability $P^N$ is zero, up to the first order in $f_s$.
It becomes nonzero only when we include contributions up to the next
higher-order terms in $f_s^2$.

As an example, we examine the case for $pp$ collisions at
$\sqrt{s_{NN}}$=200 GeV for which $f_s$=0.1, as given in Appendix A.
Equation (\ref{57}) then gives $(P^A,P^N)$=(0.043,0).  We show in
Fig.\ \ref{opps} the correlation function for two mesons with opposite
strangeness quantum numbers.  The correlation pattern of such a case
is quite distinct.  It is suppressed at $(\Delta \phi, \Delta
\eta)$$\sim$ 0 and enhanced at $(\Delta \phi \sim \pi, \Delta
\eta $$\sim$ 0).  As one notes from Eq.\ (\ref{eq69}), because $P^N$ is zero, 
the distribution $C(\Delta \phi, \Delta \eta)$ for this case is indepedent of $f_s$.

\begin{figure}[H]
\includegraphics[scale=0.75]{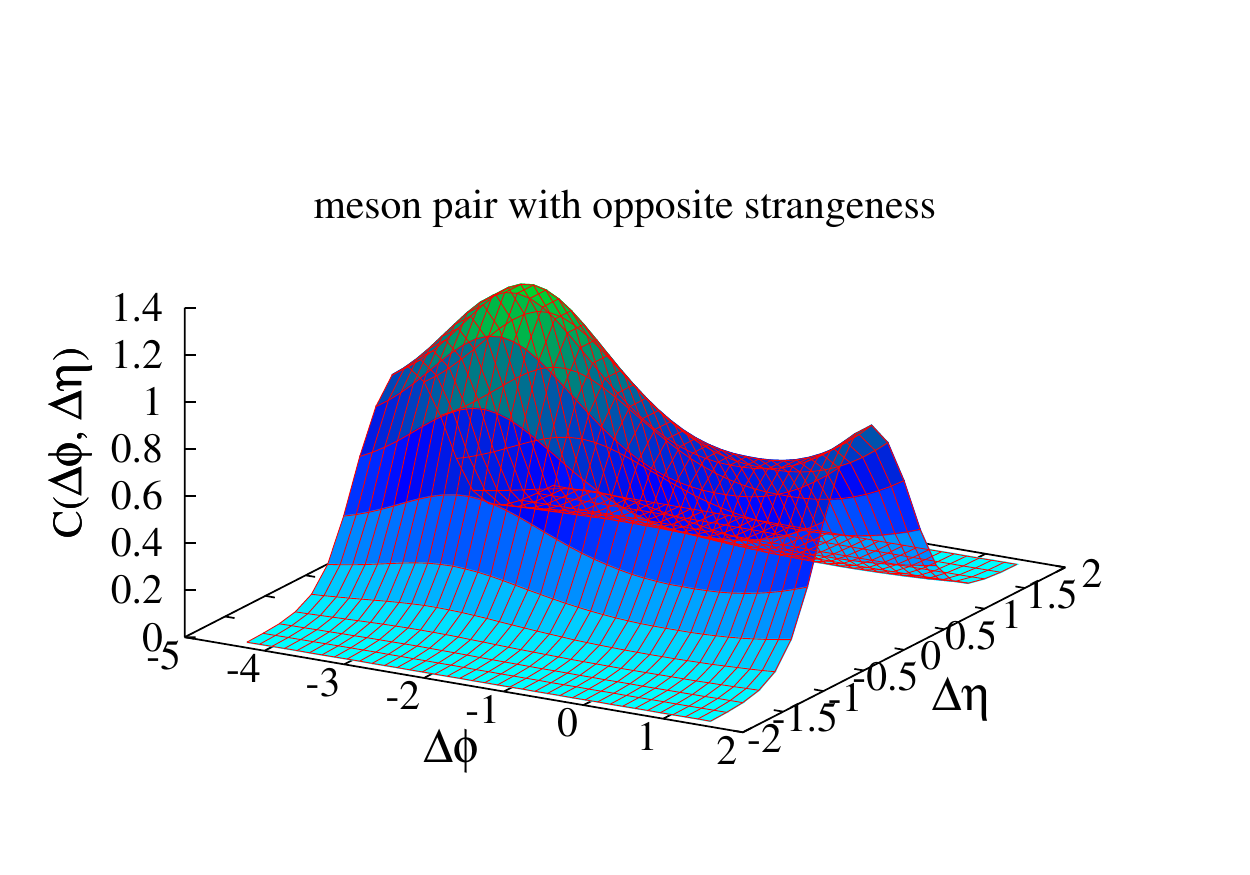}
\caption{ (Color online) The two-meson angular correlation
$C(\Delta \phi,\Delta \eta)$  for
  two mesons with opposite strangeness in a flux-tube fragmentation in $pp$ collisions.
}
\label{opps}
\end{figure}

\subsection{Production of Mesons with Other  Strangeness Configurations}

For the case of two mesons with the same nonzero strangeness, the
probabilities for adjacent and nonadjacent mesons is zero up to order
$f_s$.  They are nonzero only when we include additional
contributions from the higher order of order $f_s^2$.

From Eq.\ (\ref{59}), the case of the correlation of one meson with strangeness $|S|$=1 and
another nonstrange meson with $S=0$ gives $(P^A ,P^N)$=(0.086,0.167).  We show the correlation function 
in Fig.\ \ref{s1s0} which has a shape that is different from other configurations.
\begin{figure}[H]
\includegraphics[scale=0.75]{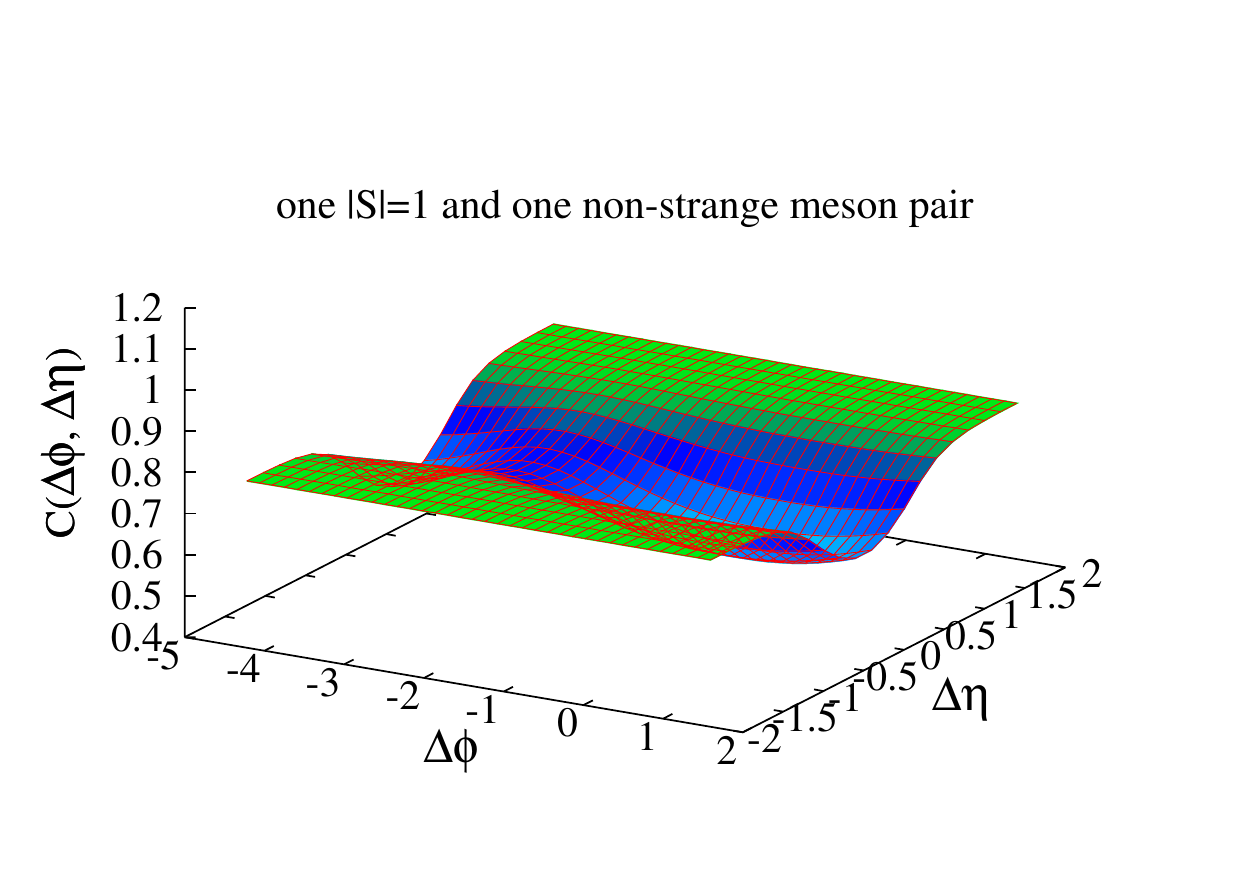}
\caption{ (Color online) The two-meson angular correlation
 $C(\Delta \phi,\Delta \eta)$  for one meson with strangeness $|S|$=1 and
another nonstrange meson with $S=0$
  in a flux-tube fragmentation in $pp$ collisions.
}
\label{s1s0}
\end{figure}

\section{Comparison with Experiment}

In a $pp$ high-energy collision, most of the produced mesons are
pions, with some fractions of strange particles and higher resonances.
The strangeness fraction is indicated by the $K^+/\pi^+$ ratio that
depends on the collision energy.  It is of order 0.1 at
$\sqrt{s_{NN}}$= 200 GeV and it decreases as the collision energy
decreases \cite{Mak15,Gaz15,Lar15,Ser15,Adakpi04,Wol11,PHO11,Vov14}.
The resonance fraction, relative to the number of pions, is of order
$\exp \{ - (m_{\rm resonance}-m_\pi)/T_{\rm eff}\}$, and it increases
with increasing collision energy.  For an effective temperature of
order 200 to 300 MeV for $pp$ collisions at $\sqrt{s_{NN}}$$\sim$ 6$-$200 GeV, the resonance fraction for $\rho$ or $\omega$ mesons is of
order $e^{-3}$ to $e^{-2}$, or 5 to 14 percent.  A resonance fraction
of such an amount would modify $P^A$ and $P^N$ by about 10 to 15\% and 
would not change greatly the gross features of the
two-hadron correlation patterns in flux-tube fragmentation shown in the
last section.

\subsection{Comparison with STAR  Two-Hadron Correlation Data 
at $\sqrt{s_{NN}}=200$ GeV}

The Star Collaboration measured the angular correlation of produced
charged hadrons represented by $\Delta \rho/\sqrt{\rho_{\rm ref}}$
which is related to the correlation function $ C(\Delta \phi,\Delta
\eta)$ of Eq.\ (\ref{eq69}) by
\begin{eqnarray}
\frac{\Delta \rho}{\sqrt{\rho_{\rm ref}}} = 
\rho_0 \{ \frac {\Delta \rho}{\rho_{\rm ref}} -1\}
=
\rho_0 \{C(\Delta \phi,\Delta \eta) -1\},
\end{eqnarray}
where $\Delta \rho$=$\rho$$-$$\rho_{\rm ref}$, $\rho$ is the
correlated distribution from sibling events, $\rho_{\rm ref}$ is the
reference distribution from mixed events, and
$\rho_0\approx\sqrt{\rho_{\rm ref}}$ is the two-dimensional  angular density
averaged over the angular acceptance $(\Delta \phi, \Delta \eta)$
\cite{STAR06twopar,Por05,Tra11,Ray11,TraKet11}.  Thus, $C(\Delta
\phi),\Delta \eta)$ has the same shape as $\Delta \rho/\sqrt{\rho_{\rm
    ref}}$ and differs by an overall constant factor and an offset.

The Star Collaboration found that if one separates the transverse
momentum regions by the boundary $p_{Tb}$=0.5 GeV/c, the correlation
$\Delta \rho/\sqrt{\rho_{\rm ref}}$ for two oppositely charged mesons
in the domain below $p_T$$<$$p_{Tb}$, as shown in Fig.\ \ref{fig1}(a),
is distinctly different from the correlation pattern in the domain
above $p_T$$>$$p_{Tb}$, as shown in Fig.\ \ref{fig1}(b).  We mentioned
earlier in the Introduction that the two-hadron correlation pattern of
Fig.\ \ref{fig1}(b) is a signature of the hard-scattering process,
with the occurrence of a jet (minijet) at $(\Delta \phi, \Delta
\eta)\sim$ 0, and another back-to-back jet at the away side with a
ridge at $\Delta\phi$$\sim$ $\pi$.

In Fig.\ \ref{fig1}(a) the small narrow peak at
$(\Delta\phi,\Delta\eta)$$\sim$0 arises from $e^+$-$e^-$ pairs
produced by photon conversion in the detector and is an experimental
artifact.  After removing the narrow sharp peak at $(\Delta \phi,
\Delta \eta)$$\sim$0 from our consideration, we find that the
theoretical correlation pattern for two oppositely charged hadrons
shown in Fig.\ \ref{opp} for a flux-tube fragmentation has the same
gross features as the experimental correlation pattern obtained by the
STAR Collaboration in Fig.\ \ref{fig1}(a), for two oppositely charged
hadrons with $p_T$$<$0.5 GeV/c in $pp$ collisions at
$\sqrt{s_{NN}}$=200 GeV.  There is a depression at $(\Delta\eta,
\Delta \phi)$$\sim$0 but an enhancement at
$(\Delta\eta$$\sim$0,$\Delta \phi$$\sim$$ \pi)$.  The similarity in
the gross features in Figs.\ \ref{fig1}(a) and \ref{opp}
indicates that in the domain with $p_T$$<$0.5 GeV/c, the dominant
particle-production mechanism is qualitatively consistent with the
flux-tube fragmentation mechanism.  Our theoretical result from a
microscopic approach supports the earlier suggestion by the STAR
Collaboration \cite{STAR06twopar,Por05,Tra11,Ray11,TraKet11} to use
the two-hadron correlation of opposite charges as the signature for
the soft particle-production process, and we identify this soft
process as flux-tube fragmentation.

There are other signatures of the flux-tube fragmentation for mesons
with charges of the same signs, or for all charged meson pairs.  The
comparison is complicated by the presence of Bose-Einstein
correlations for identical bosons.  Such a two-body Bose-Einstein
correlation is an enhanced correlation at $(\Delta \phi,\Delta
\eta)$$\sim$ 0 and it masks the large suppression of the correlation
arising from flux-tube fragmentation.  For the correlation of hadron
pairs with charges of the same sign or for all charged meson pairs, it
is however necessary to subtract the contribution from Bose-Einstein
correlations of identical bosons before one can make a meaningful
comparison.  For direct comparisons, correlation of hadrons with
opposite charges or opposite strangeness may provide the best 
signatures for flux-tube fragmentation.

\subsection{Comparison with NA61/SHINE  Two-Hadron Correlation Data 
at $\sqrt{s_{NN}}=6$$-$17 GeV}

\begin{figure}[h]
\includegraphics[width = 250 pt, height = 130 pt]{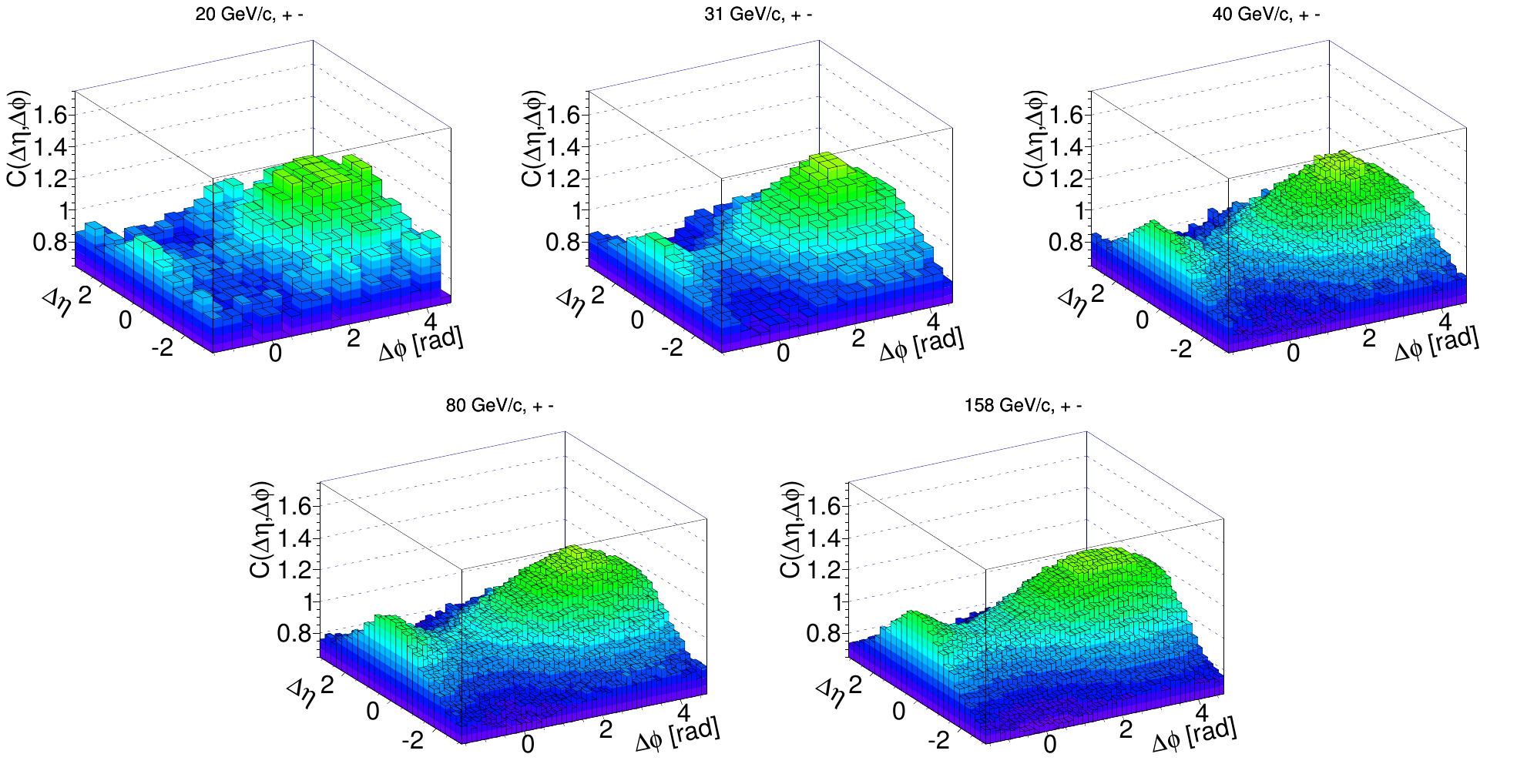}
\caption{ (Color online) NA61/SHINE two-hadron correlation data $C(\Delta \eta,\Delta \phi)$
  \cite{Mak15,Gaz15,Lar15,Ser15} for two hadrons of opposite charges
  with $p_T$$<$1.5 GeV/c, for $pp$ collisions at different energies (from
  Fig. 3 of Ref.\ \cite{Mak15}).  From left to right and from top to
  bottom, the collision energies are $\sqrt{s_{_{NN}}}$= 6.3, 7.6,
  8.7, 12.3, 17.2 GeV.
 }
\label{na61}
\end{figure} 

The NA61/SHINE Collaboration has reported the angular correlation 
$C(\Delta \eta,\Delta \phi)$\! $\propto$\! $dN/d\Delta \phi d\Delta \eta$ 
for two hadrons with opposite charges and  $p_T$$<$1.5 GeV/c, 
for $pp$ collisions at  $\sqrt{s_{NN}}$=6$-$17 GeV shown in
Fig.\ \ref{na61}  \cite{Mak15,Gaz15,Lar15,Ser15}.
One can make the following remarks:

\begin{enumerate}

\item
The experimental correlation patterns for two hadrons with opposite
charges in Fig.\ \ref{na61} have the same gross features as
the signature for the flux-tube fragmentation shown in
Fig.\ \ref{opp}, indicating that the dominant mechanism of particle
production for $p_T$ hadrons in the domain with $p_T$$<$1.5 GeV/c is
qualitatively consistent with the flux-tube fragmentation.
The pattern of the NA61/SHINE two-hadron correlation was also shown to be 
qualitatively consistent with 
the EPOS model \cite{Wer06,Mak15}.   Although  there are many different 
processes and diagrams in the EPOS model,  it is likely that the dominant process 
responsible for such a  two-hadron correlation pattern is the flux-tube fragmentation part of the EPOS model.

\item
According to Eq.\ (\ref{eq42}), the width in the $\Delta \eta$ direction
is inversely proportional to $dN/d\eta|_{\eta\sim 0}$.  Thus, as the
energy decreases, $dN/d\eta|_{\eta\sim 0}$ decreases, and one expects
the width in $\Delta \eta$ to increase.  There is a hint of such an
increase of the $\Delta \eta$ width in the data of Fig.\ \ref{na61}
but a more accurate determination of the $\Delta \eta$ width will be
needed.

\item
The correlation pattern of Fig.\ \ref{na61} remains unchanged by
shifting the $p_T$ domain up to $p_T$=1.5 GeV/c \cite{Mak15}.  This means that in the
collision energy range of $\sqrt{s_{NN}}$= 6 to 17 GeV, the domain
boundary $p_{Tb}$ between the flux-tube fragmentation process and the
hard-scattering process is greater than 1.5 GeV/c.  The boundary
$p_{Tb}$ can be located by searching for a change of the two-hadron
correlation pattern for oppositely charged hadrons from that of the flux
tube fragmentation process of Fig.\ \ref{fig1}(a) and Fig.\ \ref{opp} to that
of the hard-scattering process of Fig.\ \ref{fig1}(b), as $p_{Tb}$ increases
beyond 1.5 GeV/c.

For $\sqrt{s_{NN}}$=200 GeV, the transverse momentum boundary 
 has been located at $p_{Tb}$=0.5 GeV/c \cite{Por05}.  This shows that as the collision energy
increases, the fractional contributions from the hard-scattering
process increases, and the boundary $p_{Tb}$ has been found to move
from $p_{Tb}$ greater than 1.5 GeV/c for $\sqrt{s_{NN}}$=6$-$17 GeV
\cite{Mak15} to the lower value of $p_{Tb}$=0.5 GeV/c for
$\sqrt{s_{NN}}$=200 GeV \cite{Por05}.  The search for the boundary
$p_{Tb}$ in an energy scan will map out $p_{Tb}(\sqrt{s_{NN}})$ more precisely as a
decreasing function of the $pp$ collision energy $\sqrt{s_{NN}}$.
 
\item
There are additional predicted correlations of two hadrons with
opposite strangeness and the correlation of a strange meson with a nonstrange meson, 
as shown in Figs.\ \ref{opps} and \ref{s1s0} where the two-hadron
correlation patterns for flux-tube fragmentation are quite distinct and
may be tested with the NA61/SHINE experimental measurements.  

\item
In $pp$ collisions at NA61/SHINE energies, baryon resonances can be
produced by exciting the incident projectile and target protons.  The
decays of these baryon resonances will lead to a greater number of
$\pi^+$ as compared to $\pi^-$ and are likely to occur in the
projectile and target fragmentation regions with a diminishing
influence in the central rapidity region.  At central rapidity,
the contribution of $\pi^+$ from baryon decay has been estimated to be
about 20\% at $\sqrt{s_{_{NN}}}$=6 GeV and it decreases rapidly as a
function of collision energies \cite{Vov14}.  Correlations of two
hadrons with charges of opposite signs as shown in Fig.\ \ref{na61}
are likely to receive a greater contribution from the fragmentation of
a flux tube than from the decay of baryon resonances.

\end{enumerate}

\section{Summary and Conclusions}

The fragmentation of a flux tube is initiated by the production of
quark-antiquark pairs along the tube.  The production of the $q$-$\bar
q$ pairs occurs locally.  By the property of local conservation laws,
a quark and an antiquark produced at the same point must balance
momentum, charge, and flavor.  Subsequent to its production, a member
of the produced $q$-$\bar q$ pair interacts with its neighboring
antiparticles to form a detected meson.  Local conservation laws will
lead to correlations of the adjacently produced mesons.  On the other
hand, the production of mesons along a flux tube is ordered in
space-time and rapidity.  Mesons that are produced adjacent to each
other are also close in rapidity.  Hence, the proximity of rapidity
with a rapidity difference $\Delta y$ (or approximately a
pseudorapidity difference $\Delta \eta$) closer than a rapidity width
$w$=1$/(dN/dy)$ can signal their adjacent origin from flux-tube
fragmentation and the correlation in momentum, charge, and flavor.
Using the rapidity ordering as a signal for producing adjacent mesons,
we find that $dN/d\Delta \phi d\Delta \eta$ correlation probability or
equivalently $C(\Delta \phi,\Delta \eta)$ for two opposite charges or
strangeness mesons is suppressed on the near side at $(\Delta \phi,
\Delta \eta) \sim 0$ but the correlation probability is enhanced on
the away side at $(\Delta \phi\sim \pi, \Delta \eta \sim$ 0).  

In
addition to flux-tube fragmentation, there is the hard-scattering
process in nucleon-nucleon collisions.  The correlation patterns of
hard-scattering is well known, with a peak correlation at $(\Delta
\phi, \Delta \eta)\sim 0$ on the near-side and a ridge along the
$\Delta \eta$ direction at $\Delta \phi \sim\pi$ on the away side.
The use of the signatures for flux-tube fragmentation and hard
scattering allows one to separate out the dominance of each process in
different $p_T$ domains separated by the domain boundary $p_{Tb}$.

In our comparison of the theoretical results with experimental data,
we found that the gross features of the signature of flux-tube
fragmentation for two oppositely charged hadrons are similar to those
of the STAR \cite{STAR06twopar,Por05,Tra11,Ray11,TraKet11} and
NA61/SHINE \cite{Mak15,Gaz15,Lar15,Ser15} angular correlation data for
two hadrons with opposite charges in the low-$p_T$ region, indicating
that the dominant particle-production mechanism for low-$p_T$
particles is qualitatively consistent with flux-tube fragmentation.

We note that whereas the boundary $p_{Tb}$ that separates the flux
tube fragmentation from the hard-scattering processes in $pp$
collisions lies beyond $p_{Tb}$$>$ 1.5 GeV/c at the NA61/SHINE
collision energies of $\sqrt{s_{NN}}$=6$-$17 GeV, it shifts to
$p_{Tb}$=0.5 GeV/c at the RHIC energy of $\sqrt{s_{NN}}$=200 GeV.
Thus, the boundary $p_{Tb}$ has been observed to shift to a lower
value of $p_{Tb}$ as the collision energy increases.  Future
determination of $p_{Tb}$ in the energy scan of NA61/SHINE will map
out the boundary function $p_{Tb}(\sqrt{s_{NN}})$ as a precise
function of the collision energy $ \sqrt{s_{NN}}$.  In addition to
being an intrinsic physical property of the $pp$ collision process,
the boundary function $p_{Tb} (\sqrt{s_{NN}})$ separating the two
processes in $pp$ collisions also may have implications on the early
evolution and thermalization of particles, the quenching
of jets, and the formation of the quark-gluon plasma, in high-energy
nucleus-nucleus collisions.

Returning to the decomposition of particle-production mechanisms into
soft and hard components in $pp$ collisions, our theoretical result
from a microscopic approach of the flux-tube fragmentation supports
the earlier phenomenological suggestion by the STAR Collaboration in
Refs.\ \cite{STAR06twopar,Por05,Tra11,Ray11,TraKet11} to associate the
correlation of two hadrons with opposite charges to the dominance of
the soft component of particle production in the domain with $p_T$$<$0.5
GeV/c at $\sqrt{s_{NN}}$=200 GeV.  However, our result from a microscopic approach does not support
the second phenomenological suggestion in
Refs.\ \cite{STAR06twopar,Por05,Tra11,Ray11,TraKet11} of a one-dimensional
Gaussian of $\Delta \eta$ independent of $\Delta \phi$.  For a better
accuracy, a distribution such as Eq.\ (\ref{A}) would be a better
representation of the two-hadron correlation function of the `soft'
component for hadrons with opposite charges than a one-dimensional
Gaussian.

The interference of the Bose-Einstein correlation makes it difficult
to use the correlations of two hadron with charges of the same sign or two charged
hadrons as signatures of the flux-tube fragmentation process.  In
addition to the the Bose-Einstein correlation of identical bosons,
there is also the production of resonances that will complicate the
pattern of two-hadron correlation.  The untangling of these many
different correlations patterns as a function of collision energies
will provide more precise information on the changing role of the flux-tube
fragmentation and hard scattering as a function of the collision
energy.

Our successes in identifying the signature of the fragmentation of a
flux tube will provide a window to examine further the dynamics of the
many-meson system and the associated many-meson correlations produced in
the flux-tube fragmentation.  The space-time-rapidity ordering
stipulates a very simple many-meson space-time-rapidity correlations.
One may wish to find out how a chain of many mesons correlate with
each other in their azimuthal angles and rapidities, after their
production in a flux-tube fragmentation.  An interesting experimental
and theoretical question is the case of an exclusive measurement on an
event-by-event basis if the momenta of all mesons in the flux-tube
fragmentation have been measured, and whether it is possible to piece
together different produced mesons in that event to come up with the
configuration of the whole system at the moment of its fragmentation.
These and many other questions may be opened up for examination upon a
successful experimental search for the signature of the flux-tube
fragmentation.

\appendix

\section{ Relation between  $K^+/\pi^+$ ratio and $f_s$ }

In $pp$ collisions at $\sqrt{s_{NN}}$$\sim$ 6$-$17 GeV, the ratios
$K^+/\pi^+$, $K^-/\pi^-$, and $\pi^+/\pi^-$ depend on the collision
energy and are affected by the rescattering and decays of baryonic
resonances formed by the inelastic scattering of the incident protons.
As a consequence, the ratio $K^+/\pi^+$ is greater than $K^-/\pi^-$
\cite{Mak15,Gaz15,Lar15,Vov14}.  In higher energy $pp$ collisions at
$\sqrt{s_{NN}}$= 200 GeV, the effects of rescattering and the decays
of baryonic resonances are less important in the central rapidity
region, and $K^+/\pi^+$$\sim$$K^-/\pi^-$$\sim$ 0.10$-$0.12, and
$\pi^+/\pi^-$$\sim$1 \cite{Adakpi04}.  It becomes appropriate to
estimate the strangeness suppression factor $f_s$ at $\sqrt{s_{NN}}$=
200 GeV using the flux-tube fragmentation model.  In such a model,
Table V gives the number of cases $N^{A(ud,s)}(Q_{12},Q_{34})$
with meson charges $(Q_{12},Q_{34})$ in various charge states in the
nonstrange sector, and the strange sector, for a pair of adjacent
mesons.  We can treat the pair production at all other vertices as a
copy of Table V, which can then be used to calculate the ratio
of the single-particle production of $\pi^+$ and $K^+$.  We find
from Table V the number of cases of nonstrange $(u\bar d)$ mesons with a
positive charge,
\begin{eqnarray}
 \sum_{Q_{12},Q_{34}=+1,(u\bar d) }N^{(ud)}(Q_{12},Q_{34})
= 4+2f_s.
\end{eqnarray} 
We find from table V the number of cases of strange $(u\bar s)$ mesons with 
a positive charge,
\begin{eqnarray}
\sum_{Q_{12},Q_{34}=+1,(u\bar s) }N^{(s)}(Q_{12},Q_{34})
=4f_s.  
\end{eqnarray} 
Therefore, we have
\begin{eqnarray}
&&\!\!\frac{\text{(number of positive strange} ~ K^+\text{ meson)}}{\text{(number of positive nonstrange ~}\pi^+~\text{mesons } )} \sim \frac{ (u\bar s)}{ (u\bar d)}
\nonumber\\
&&
\hspace*{5.5cm}
=\frac{4f_s}{4+2f_s}.
\end{eqnarray}
Experimentally, at $\sqrt{s}>$200 GeV, the STAR Collaboration gives \cite{Adakpi04}
\begin{eqnarray}
\frac{K^+}{\pi^+}=\frac{\text{(number of } K^+)}{\text{(number of } \pi^+)}=0.10~ {\rm to}~0.12,
\end{eqnarray}
which leads to the estimate  
\begin{eqnarray}
f_s=0.105~{\rm to~} 0.126.
\end{eqnarray}
Although Table V may appear complicated, it is intuitively reasonable
that $K^+/\pi^+$ should be nearly equal to $f_s$ in flux-tube
fragmentaion because given a $u$ quark, $\pi^+$ is the result of a
$d$-$\bar d$ pair production, and $K^+$ is the result of an $s$-$\bar
s$ pair production with a reduced probability $f_s$.  Hence,
$K^+$/$\pi^+$ should be nearly equal to $f_s$.

It should be pointed out that the strangeness suppression factor we
have estimated is significantly smaller than the value of $f_s$=0.3
used in the standard Lund Monte Carlo programs \cite{Sj87,Sjo14}.  The
determination of $f_s$ in the flux-tube fragmentation from the
$K^+/\pi^+$ ratio is best carried out when the flux-tube fragmentation
process can be isolated without interference from contributions from
other mechanisms.  However, for $pp$ collisions at energies much below
$\sqrt{s}=200$ GeV, the production is influenced by contributions from
baryon resonance production in which an incident proton is excited to
a baryon resonance which subsequently emit a meson \cite{Gaz15,Lar15}.
The baryon resonances are produced mostly at the projectile and target
fragmentation regions but have contributions at central rapidity when
the collision energies are low as in the $\sqrt{s}\sim $10 GeV region.
The determination of $f_s$ in these lower-energy region will be
affected by the parameters specifying baryon resonance production
processes. Only for high enough energies at central rapidity can the
flux-tube fragmentation processes be isolated with small contributions
from baryon resonance contributions.  Future extraction of $f_s$ for
the flux-tube fragmentation around $\sqrt{s}=200$ GeV with $p_T<0.5$
GeV/c will be of interest in clarifying the magnitude of the
strangeness suppression factor $f_s$.

\section{Charge and Strangeness Configurations for Flux-Tube Fragmentation of Two nonadjacent Mesons with Three Flavors}

We examine the quark charge and flavor configurations in two
nonadjacent mesons $P_{12}$ and $P_{34}$ in flux-tube fragmentation
with three flavors as depicted in Fig.\ \ref{fig3}.  The constituents
$p_2$ and $p_3$ are produced at different vertices and are not
constrained by charge and flavor conservation. The cases of possible
charge and strangeness configurations come in three parts.  By
including only cases up to the first order in $f_s$, we list the
charge and strangeness configurations in which $p_2$ is a $u$ quark in
Table \ref{t10} as Part I, $p_2$ is a $d$ quark in Table \ref{t13} as
Part II, and $p_2$ is an $s$ quark in Table \ref{t16} as Part III.
The corresponding number of cases $N^{N(ud)}$ in the
nonstrange sector and $N^{N(s)}$ in the strange sector
are given in Tables \ref{t11}, \ref{t12}, \ref{t14}, \ref{t15},
\ref{t17}, and \ref{t18}.

In a flux-tube fragmentation with three flavors for nonadjacent
mesons $P_{12}$ and $P_{34}$ in which $p_2$ is a $u$ quark, we can use
Table \ref{t10} to construct Table \ref{t11} as Part I of
the number of cases $N^{N(ud,s)}(Q_{12},Q_{34})$ in the strange and
nonstrange sectors.

\begin{table}[H]
\centering
\label{t1}
\caption { 
Part I of quark and antiquark configurations with three flavors for
two nonadjacent mesons $P_{12}$ and $P_{34}$ in a flux-tube
fragmentation with three flavors in which $p_2$ is a $u$ quark.
}
\label{t10}
\vspace*{0.2cm}
\begin{tabular}{|c|c|c|c|c|c|c|c|c|}
\cline{1-9}
     $p_1$    &  $p_2$  &  $Q_{12}$  & $S_{12}$  &$p_3$  &  $ p_4$  &  $Q_{34}$   &  $S_{34}$
& Order
  \\
\cline{1-9}
$\bar u$  & $u$ &  0 & 0 &  $\bar u$  & $d$ & -1  & 0  & 1
\\
$\bar d$  &$u$ &  1 &  0 &  $\bar u$  & $d$ & -1 & 0 & 1
\\ 
$\bar u$  & $u$ &  0 &  0 &  $\bar u$  & $u$ & 0 & 0 & 1
\\ 
$\bar d$  &$u$ &  1 &  0 &  $\bar u$  & $u$ & 0 & 0 & 1
\\
$\bar u$  & $u$ &  0 &  0 &  $\bar d$  & $d$ & 0 & 0 & 1
\\ 
$\bar d$  & $u$ & 1 &  0 &  $\bar d$  & $d$ & 0 & 0 & 1
\\
$\bar u$  & $u$ &  0 &  0 &  $\bar d$  & $u$ & 1 & 0 & 1
\\ 
$\bar d$  &$u$ &  1 & 0 &   $\bar d$  & $u$ & 1 & 0 & 1
\\
\hline
$\bar s$  &$u$ &  1 &  1 &  $\bar u$  & $d$ & -1 & 0 & $f_s$ 
\\
$\bar s$  &$u$ &  1 & 1 & $\bar u$  & $u$ & 0 & 0 & $f_s$ 
\\
$\bar u$  & $u$ &  0 &  0 &  $\bar u$  & $s$ & -1 & -1& $f_s$ 
\\
$\bar d$  &$u$ &  1 &  0 &  $\bar u$  & $s$ & -1 & -1&  $f_s$ 

\\
$\bar s$  & $u$ & 1 & 1 & $\bar d$  & $d$ & 0 & 0& $f_s$
\\
$\bar s$  &$u$ &  1 & 1 & $\bar d$  & $u$ & 1& 0 &$f_s$
\\ 
$\bar u$  & $u$ &  0 &  0 &  $\bar d$  & $s$ & 0 & -1&$f_s$
\\ 
$\bar d$  & $u$ & 1 &  0 &  $\bar d$  & $s$ & 0 & -1&$f_s$
\\
$\bar u$  & $u$ &  0 &  0 &   $\bar s$  & $d$ & 0 & 1&$f_s$
\\ 
$\bar d$  & $u$ & 1 & 0 &   $\bar s$  & $d$ & 0 & 1&$f_s$
\\
$\bar u$  & $u$ &  0 &  0 &  $\bar s$  & $u$ & 1 &  1&$f_s$
\\ 
$\bar d$  &$u$ &  1 &  0 &  $\bar s$  & $u$ & 1 &  1&$f_s$
\\
\hline
\end{tabular}
\end{table}

\begin{table}[H]
\centering 
\caption { Part I of the number of cases $N^{N(ud,s)}(Q_{12},Q_{34})$
  for nonadjacent mesons in different charge states $(Q_{12},Q_{34})$
  in the nonstrange sector $(ud)$ and the strange sector $(s)$, in
  the fragmentation of a flux tube with three flavors in which $p_2$
  is an $u$ quark.
}
\label{t11}
\vspace*{0.2cm}
\begin{tabular}{|c|c|c|c|c|}
\cline{3-5}
  \multicolumn{1}{c}       {}    &  &  $Q_{34}$=-1  &  $Q_{34}=0$  &  $Q_{34}$=1 \\ 
\cline{1-5}
 &$Q_{12}$=-1  &    0    &   0  &   0  
\\
$N^{N(ud)}(Q_{12},Q_{34})$  & $Q_{12}$=~0  &  1      &     2   & 1
\\ 
  & $Q_{12}$=~1  &    1  &     2    &  1
\\ \hline
\cline{1-5}
 &$Q_{12}$=-1  &    0    &   0   &     0
\\
$N^{N(s)}(Q_{12},Q_{34})$ & $Q_{12}$=~0  &  $f_s$     &     2$f_s$   &  $f_s$ 
\\ 
 & $Q_{12}$=~1  &   2$f_s$   &    4 $f_s$     & 2$f_s$ 
\\ \hline
\end{tabular}
\end{table}

From Table \ref{t10} we can similarly construct Table
\ref{t12} as Part I of the number of cases
$N^{N(ud,s)}(S_{12},S_{34})$ for nonadjacent mesons $P_{12}$ and
$P_{34}$ in different strangeness states $(S_{12},S_{34})$, in the
fragmentation of a flux tube with three flavors in which $p_2$ is a
$u$ quark.

\begin{table}[H]
\centering
\caption { Part I of the number of cases $N^{N(s)}(S_{12},S_{34})$ for
  nonadjacent mesons $P_{12}$ and $P_{34}$ in different strangeness
  states $(S_{12},S_{34})$ in a flux-tube fragmentation with three
  flavors, in which $p_2$ is a $u$ quark.
}
\label{t12}
\vspace*{0.2cm}
\begin{tabular}{|c|c|c|c|c|}
\cline{3-5}
  \multicolumn{1}{c}       {}    &  &  $S_{34}$=-1  &  $S_{34}=0$  &  $S_{34}$=1 \\ 
\cline{1-5}
&$S_{12}$=-1  &    0    &   0  &    0
\\
 $N^{N(s)}(S_{12},S_{34})$ & $S_{12}$=~0  &  $4f_s$     &     0  &  $4 f_s$
\\ 
  & $Q_{12}$=~1  &   0  &   4$ f_s$     & 0
\\ \hline
\end{tabular}
\end{table}

In a flux-tube fragmentation with three flavors for nonadjacent
mesons $P_{12}$ and $P_{34}$ in which $p_2$ is a $d$ quark, the
possible charge and strangeness configurations are listed in Table
\ref{t13} as Part II.

\begin{table}[H]
\centering
\caption {
Part II of quark and antiquark configurations with three flavors for
two nonadjacent mesons $P_{12}$ and $P_{34}$ in a flux-tube
fragmentation with three flavors in which $p_2$ is a $d$ quark.
}
\label{t13}
\vspace*{0.2cm}
\begin{tabular}{|c|c|c|c|c|c|c|c|c|}
\cline{1-9}
     $p_1$    &  $p_2$  &  $Q_{12}$   &  $s_{12}$ & $p_3$  &  $ p_4$  &  $Q_{34}$  &  $s_{34}$ 
& Order
  \\
\cline{1-9}
$\bar u$  & $d$ &  -1 & 0 &  $\bar u$  & $d$ & -1 & 0 & 1
\\
$\bar d$  &$d$ &  0 & 0& $\bar u$  & $d$ & -1 & 0 & 1
\\ 
$\bar u$  & $d$ & -1 &  0 & $\bar u$  & $u$ & 0 & 0& 1
\\ 
$\bar d$  &$d$ & 0&  0 & $\bar u$  & $u$ & 0 & 0& 1

\\ 
$\bar u$  & $d$ &  -1 & 0& $\bar d$  & $d$ & 0 & 0& 1
\\ 
$\bar d$  & $d$ & 0 & 0& $\bar d$  & $d$ & 0 & 0& 1
\\
$\bar u$  & $d$ &  -1 & 0& $\bar d$  & $u$ & 1 & 0& 1
\\ 
$\bar d$  &$d$ &  0 & 0& $\bar d$  & $u$ & 1 & 0& 1
\\ 
\hline
$\bar s$  &$d$ &  0 & 1&  $\bar u$  & $d$ & -1 & 0 & $f_s$ 
\\
$\bar s$  &$d$ &  0 & 1& $\bar u$  & $u$ & 0 & 0& $f_s$ 
\\
$\bar u$  & $d$ & -1 &  0  & $\bar u$  & $s$ & -1 & -1  & $f_s$
\\
$\bar d$  &$d$ &  0 & 0& $\bar u$  & $s$ & -1 & -1  & $f_s$
\\
$\bar s$  & $d$ & 0 & 1& $\bar d$  & $d$ & 0 & 0 & $f_s$

\\ 
$\bar s$  &$d$ &  0 & 1& $\bar d$  & $u$ & 1& 0 & $f_s$
\\ 
$\bar u$  & $d$ &  -1 & 0& $\bar d$  & $s$ & 0 & -1  & $f_s$
\\ 
$\bar d$  & $d$ & 0 & 0& $\bar d$  & $s$ & 0 & -1  & $f_s$
\\
$\bar u$  & $d$ &  -1 & 0 & $\bar s$  & $d$ & 0 & 1  & $f_s$
\\ 
$\bar d$  & $d$ & 0 & 0& $\bar s$  & $d$ & 0 & 1 & $f_s$
\\
$\bar u$  & $d$ &  -1 & 0& $\bar s$  & $u$ & 1 & 1 & $f_s$
\\ 
$\bar d$  &$d$ &  0 & 0& $\bar s$  & $u$ & 1 & 1  & $f_s$
\\\hline
\end{tabular}
\end{table}

From Table \ref{t13} we can construct Table \ref{t14} as
Part II of the number of cases $N^{N(ud,s)}(Q_{12},Q_{34})$ for
nonadjacent mesons $P_{12}$ and $P_{34}$ to be in different charge
states $(Q_{12},Q_{34})$ in the nonstrange and strange sectors, in a
flux-tube fragmentation with three flavors in which $p_2$ is a $d$
quark.

\begin{table}[H]
\centering
\caption { 
Part II of the number of cases $N^{N(ud,s)}(Q_{12},Q_{34})$ for
nonadjacent mesons in different charge states $(Q_{12},Q_{34})$ in
the nonstrange sector $(ud)$ and the strange sector $(s)$, in a flux
tube fragmentation with three flavors in which $p_2$ is a $d$ quark.
}
\label{t14}
\vspace*{0.2cm}
\begin{tabular}{|c|c|c|c|c|}
\cline{3-5}
  \multicolumn{1}{c}       {}    &  &  $Q_{34}$=-1  &  $Q_{34}=0$  &  $Q_{34}$=1 \\ 
\cline{1-5}
 &$Q_{12}$=-1  &    1    &   2  &   1  
\\
 $N^{N(ud)}(Q_{12},Q_{34})$ & $Q_{12}$=~0  &  1      &     2   & 1
\\ 
  & $Q_{12}$=~1  &    0  &     0    &  0
\\ \hline
\cline{1-5}
 &$Q_{12}$=-1  &    $f_s$     &   2$f_s$    &     $f_s$ 
\\
 $N^{N(s)}(Q_{12},Q_{34})$ & $Q_{12}$=~0  &  2$f_s$      &     4$f_s$    &  2$f_s$ 
\\ 
 & $Q_{12}$=~1  &   0  &    0     & 0
\\ \hline
\end{tabular}
\end{table}

From Table \ref{t13} we can similarly construct Table
\ref{t15} as Part II of the number of cases
$N^{N(ud,s)}(S_{12},S_{34})$ for nonadjacent mesons $P_{12}$ and
$P_{34}$ in different strangeness states $(S_{12},S_{34})$, in a flux
tube fragmentation with three flavors in which $p_2$ is a $d$ quark.

\begin{table}[H]
\caption { 
Part II of the number of cases $N^{N(s)}(S_{12},S_{34})$ for
nonadjacent mesons $P_{12}$ and $P_{34}$ in different strangeness
states $(S_{12},S_{34})$, in a flux-tube fragmentation with three
flavors in which $p_2$ is a $d$ quark.
}
\label{t15}
\vspace*{0.2cm}
\begin{tabular}{|c|c|c|c|c|}
\cline{3-5}
  \multicolumn{1}{c}       {}    &  &  $S_{34}$=-1  &  $S_{34}=0$  &  $S_{34}$=1 \\ 
\cline{1-5}
&$S_{12}$=-1  &    0    &   0  &    0
\\
 $N^{A(s)}(S_{12},S_{34})$ & $S_{12}$=~0  &  $4f_s$     &     0  &  $4 f_s$
\\ 
  & $S_{12}$=~1  &   0  &   4$ f_s$     & 0
\\ \hline
\end{tabular}
\end{table}

In a flux-tube fragmentation with three flavors in which $p_2$ is an
$s$ quark, we list all possible charge and strangeness configurations
for nonadjacent mesons $P_{12}$ and $P_{34}$ in Table \ref{t16} as
Part III.

\begin{table}[H]
\centering
\caption { 
Part III of quark and antiquark configurations for two nonadjacent
mesons $P_{12}$ and $P_{34}$ in a flux-tube fragmentation with three
flavors in which $p_2$ is an $s$ quark.
 }
\label{t16}
\vspace*{0.3cm}
\begin{tabular}{|c|c|c|c|c|c|c|c|c|}
\cline{1-9}
     $p_1$    &  $p_2$  &  $Q_{12}$  &  $S_{12}$  & $p_3$  &  $ p_4$  &  $Q_{34}$ &  $S_{34}$ 
& Order 
  \\
\cline{1-9}
$\bar u$  & $s$ &  -1 & -1 & $\bar u$  & $d$ & -1 & 0 & $f_s$
\\
$\bar d$  &$s$ &  0  & -1& $\bar u$  & $d$ & -1 & 0 & $f_s$
\\ 
$\bar u$  & $s$ &  -1  & -1& $\bar u$  & $u$ & 0 & 0 & $f_s$
\\ 
$\bar d$  &$s$ &  0  & -1& $\bar u$  & $u$ & 0 & 0& $f_s$
\\  
$\bar u$  & $s$ &  -1  & -1&  $\bar d$  & $d$ & 0 & 0& $f_s$
\\ 
$\bar d$  & $s$ & 0  & -1& $\bar d$  & $d$ & 0 & 0& $f_s$
\\
$\bar u$  & $s$ &  -1  & -1& $\bar d$  & $u$ & 1 & 0& $f_s$
\\ 
$\bar d$  &$s$ &  0  & -1 & $\bar d$  & $u$ & 1 & 0& $f_s$
\\\hline
\end{tabular}
\end{table}

From  Table \ref{t16} we can construct Table \ref{t17} as
Part III of the number of cases $N^{N(ud,s)}(Q_{12},Q_{34})$ for
nonadjacent mesons $P_{12}$ and $P_{34}$ to be in different charge
states $(Q_{12},Q_{34})$ in the strange and nonstrange sectors, in a
flux-tube fragmentation with three flavors in which $p_2$ is an $s$
quark.

\begin{table}[h]
\centering
\caption { 
Part III of the number of cases $N^{N(ud,s)}(Q_{12},Q_{34})$ for
nonadjacent mesons in different charge states $(Q_{12},Q_{34})$ in
the nonstrange sector $(ud)$ and the strange sector $(s)$, in a flux
tube fragmentation with three flavors in which $p_2$ is an $s$ quark.
 }
\label{t17}
\vspace*{0.2cm}
\begin{tabular}{|c|c|c|c|c|}
\cline{3-5}
  \multicolumn{1}{c}       {}    &  &  $Q_{34}$=-1  &  $Q_{34}=0$  &  $Q_{34}$=1 \\ 
\cline{1-5}
 &$Q_{12}$=-1  &    0    &   0  &   0  
\\
  $N^{N(ud)}(Q_{12},Q_{34})$ & $Q_{12}$=~0  &  0      &     0   & 0
\\ 
 & $Q_{12}$=~1  &    0  &     0    &  0
\\ \hline
\cline{1-5}
&$Q_{12}$=-1  &    $f_s$     &   2$f_s$    &    $f_s$ 
\\
$N^{N(s)}(Q_{12},Q_{34})$  & $Q_{12}$=~0  &  $f_s$      &     2$f_s$    &  $f_s$ 
\\ 
 & $Q_{12}$=~1  &   0  &    0     & 0
\\ \hline
\end{tabular}
\end{table}

From the above Table \ref{t16} we can similarly construct Table
\ref{t18} as Part III of the number of cases
$N^{N(ud,s)}(S_{12},S_{34})$ for nonadjacent mesons $P_{12}$ and
$P_{34}$ in different strangeness states $(S_{12},S_{34})$, in a flux
tube fragmentation with three flavors in which $p_2$ is an $s$ quark.

\begin{table}[h]
\centering
\caption { 
Part III of the number of cases $N^{N(s)}(S_{12},S_{34})$ for
nonadjacent mesons $P_{12}$ and $P_{34}$ in different strangeness
states $(S_{12},S_{34})$ in a flux-tube fragmentation, in a flux-tube
fragmentation with three flavors in which $p_2$ is an $s$ quark.
}
\label{t18}
\vspace*{0.2cm}
\begin{tabular}{|c|c|c|c|c|}
\cline{3-5}
  \multicolumn{1}{c}       {}    &  &  $S_{34}$=-1  &  $S_{34}=0$  &  $S_{34}$=1 \\ 
\cline{1-5}
&$S_{12}$=-1  &    0   &   $8f_s$  &    0
\\
 $N^{N(s)}(S_{12},S_{34})$  & $S_{12}$=~0  &  0    &     0  &  0
\\ 
  & $Q_{12}$=~1  &   0  &   0     & 0
\\ \hline
\end{tabular}
\end{table}

Upon adding the contributions from all contributions from Parts I, II
and III, we obtain Tables \ref{t8} and \ref{t9} in Section VII.

\vspace*{0.3cm}
\centerline{\bf Acknowledgments}
\vspace*{0.3cm}

The author would like to thank Drs.  Elena Kokoulina and G. Feofilov 
for helpful
discussions. The research was supported in part by the Division of
Nuclear Physics, U.S. Department of Energy under Contract
DE-AC05-00OR22725.


\begin{thebibliography}{99}


\bibitem{Sch51}
J. Schwinger, Phys. Rev. {\bf 82}, 664 (1951). 

\bibitem{Sch62}
J. Schwinger, Phys. Rev. {\bf 128}, 2425 (1962). 


\bibitem{Nam70} 
Y. Nambu, Lectures at Copenhagen Symposium (1970).

\bibitem{Bjo73}
J. D. Bjorken, Lectures presented at the 1973  Proceedings  of 
the Summer Institute on  Particle  Physics,  edited  by  Zipt,     
SLAC-167 (1973).




\bibitem{Cas74}
A. Casher, J. Kogut, and  L.  Susskind,  Phys.  Rev.  D10,  732 
(1974).

\bibitem{Art74} 
X. Artru and G. Mennessier, Nucl. Phys. B70, 93 (1974).


\bibitem{And79}
B. Andersson, G.  Gustafson  and  C. Peterson, Z. Phys. C1, 105 (1979);
B. Andersson, G.  Gustafson  and  B. S\"oderberg, Z. Phys. C20, 317 (1983).


\bibitem{And83}
A comprehensive review of the application of the Flux-Tube 
Fragmentation Model for nucleon-nucleon and 
$e^+-e^-$ 
collisions can be found in 
B. Andersson, G. Gustafson,  G.  Ingelman,  and  T.  Sj\"ostrand, 
Phys. Rep. 97, 31 (1983), and X. Artru, Phys. Rep. 97, 147 (1983).


\bibitem{And83a} B. Andersson, G. Gustafson, and T. Sj\"ostrand,
  Zeit. f{\"u}r Phys. {\bf C20}, 317 (1983); T. Sj\"ostrand and M. Bengtsson, Computer Physics
  Comm.  {\bf 43}, 367 (1987); B. Andersson, G. Gustavson, and
  B. Nilsson-Alqvist, Nucl. Phys. {\bf B281}, 289 (1987).

\bibitem{Sjo14}
T. Sj\" ostrand $et~al.$, 
    { An Introduction to PYTHIA 8.2},  Comput. Phys. Commun. {\bf 191},159 (2015), arXiv:1410.3012 .

\bibitem{Art84}
X. Artru, 
Z. Phys. C26, 83 (1984).


\bibitem{Sjo86}
T. Sj\"ostrand, Comp. Phys. Comm. 39, 347 (1986);
T. Sj\"ostrand, and M. Bengtsson, Comp. Phys. Comm. 43, 367 (1987).

\bibitem{Wan88}
R. C. Wang and C. Y. Wong,
Phys. Rev. {\bf D38}, 2890 (1988).

\bibitem{Pav91}
H-P. Pavel and D. Brink, Zeit. Phys. {\bf C51}, 119 (1991).

\bibitem{Won91a}
C. Y. Wong, R. C. Wang, and C. C. Shih, Phys. Rev. {\bf D44}, 257  (1991).

\bibitem{Won91b}
C. Y. Wong, R. C. Wang, Phys. Rev. {\bf D44}, 679 (1991).

\bibitem{Gat92}
G. Gatoff and C.Y. Wong,
Phys. Rev. {\bf D 46}, 997 (1992);
and C.Y. Wong and G. Gatoff, Phys. Rep. {\bf 242},  1994, 489 (1994).



\bibitem{Won95}
C. Y. Wong, R.C Wang, and  J.S. Wu, Phys. Rev {\b  D51}, 3940 (1995).

\bibitem{Feo08}
D.A. Derkach, G.A. Feofilov, Phys. Atom. Nucl. {\bf 71}, 2087  (2008);
E.O. Bodnya, V.N. Kovalenko, A.M. Puchkov, G.A. Feofilov,  AIP
 Conf.Proc. {\bf 1606},  273 (2014); Adam, Jaroslav et al.)
JHEP 05 (2015);
G. Feofilov,
I. Altsybeev,  O. Kochebina, XXII International Baldin Seminar on High Energy
Physics Problems 15-20 September, 2014 JINR, Dubna, Russia, PoS(Baldin
ISHEPP XXII) 067. 


\bibitem{Won94} C. Y. Wong, {\it Introduction to High-Energy Heavy-Ion
  Collisions}, World Scientific Publisher, Singapore, 1994.



\bibitem{Bla74}  R.~Blankenbecler and S.~J.~Brodsky, Phys.\ Rev.\  D {\bf  10},  2973 (1974);
                 R.~Blankenbecler, S.~J.~Brodsky and  J.~Gunion, Phys.\ Rev.\ D  {\bf 12},
                 3469 (1975);  E.~A.~Schmidt and R.~Blankenbecler,  Phys.\ Rev.\ D  {\bf 15},
                 332 (1977).

\bibitem{Ang78} A.L.S. Angelis $et~al.$ (CCOR Collaboration),  Phys.\ Lett.\ B {\bf 79}, 505 (1978).

\bibitem{Ang79}
A.L.S. Angelis $et~al.$,(CCOR Collaboration),  Physica Scripta {\bf 19}, 116 (1979);
A.L.S. Angelis $et~al.$, (CCOR Collaboration), Phys. Lett.  {\bf B97}, 163 (1980).


\bibitem{Fey78} R. P. Feynman, R. D. Field and G. C. Fox,  Phys.\ Rev.\ D {\bf 18}, 3320 (1978).

\bibitem{Owe78} J. F. Owens, E. Reya, and M. Gl\" ck,  Phys.\ Rev.\ D {\bf 18}, 1501 (1978).

\bibitem{Duk84} D.~W.~Duke, J.~F.~Owens, Phy.\ Rev.\  D {\bf 30}, 49 (1984).

\bibitem{Sj87} T.~Sj\"ostrand and M.~ van Zijl, Phys.\ Rev.\ D  {\bf 36}, 2019 (1987);
               R.~Corke and T.~Sj\" ostrand, JHEP {\bf 1103},
               032 (2011); T. Sj\" ostrand and P. Z. Skands,
               Eur.\ Phys.\ J.\ C {\bf 39}, 129 (2005);
               T. Sj\" ostrand and P. Z. Skands, JHEP {\bf 03}, 053 (2004).

\bibitem{UA188} C. Albajar  $et~al.$  (UA1 Collaboration),  Nucl.\ Phys.\ B {\bf 309},  405 (1988).

\bibitem{Wan91} X.N. Wang and M. Gyulassy, Phys.\ Rev.\ D {\bf 44}, 3501 (1991).

\bibitem{Rak13} J.~Rak and M.~J.~Tannenbaum, {\it High-$p_T$ Physics in the Heavy Ion },
             Cambridge University Press, Cambridge, 2013.

\bibitem{Arl10} F.~Arleo, S.~Brodsky, D.~S.~Hwang and A.~M.~Sickles, Phys.\ Rev.\
                Lett. {\bf 105}, 062002 (2010).

\bibitem{Won12} C.~Y.~Wong and G.~Wilk, Acta\ Phys.\ Pol.\ B {\bf 43}, 2047 (2012).


\bibitem{Won13} C.~Y.~Wong and G.~Wilk, Phys.\ Rev.\ D  {\bf 87}, 114007 (2013).


\bibitem{Won13a} C.~Y.~Wong, G.~Wilk,  arXiv:1309.7330.


\bibitem{Won14EPJ} C. Y. Wong, G. Wilk, L. J. L. Cirto and C. Tsallis;   EPJ Web Conf.
                 {\bf 90},  04002 (2015),  arXiv:1412.0474.

\bibitem{CTWW14} L.~J.~L.~Cirto, C.~Tsallis, C.-Y.~Wong and G.~Wilk,  arXiv:1409.3278.

\bibitem{T1} C.~Tsallis, J.\ Stat.\ Phys.\ {\bf 52}, 479 (1988) and Eur.\ Phys.\
             J.\ A {\bf 40}, 257 (2009);   M.\ Gell-Mann and C.\ Tsallis eds.,
            \emph{Nonextensive Entropy -- Interdisciplinary Applications}
            (Oxford University Press, New York, 2004). 

\bibitem{Won15} C. Y. Wong, G. Wilk, L. J. L. Cirto and C. Tsallis, 
Phys. Rev. {\bf D91}, 114027 (2015). 

\bibitem{H} R.~Hagedorn, Riv.\ Nuovo\ Cimento\ {\bf 6},  1  (1984).


\bibitem{Michael} C.~Michael and L.~Vanryckeghem, J.\ Phys.\ G {\bf 3}, L151 (1977);
                  C.~Michael, Prog.\ Part.\ Nucl.\ Phys. {\bf 2}, 1 (1979).




\bibitem{STAR06twopar} J. Adams $et~al.$ (STAR Collaboration), Phys.\ Rev.\ D {\bf 74}, 032006 (2006).


\bibitem{Por05} R. J. Porter and T. A. Trainor,  (STAR Collaboration), J.\ Phys.\ Conf.\ Ser.\
                {\bf 27}, 98  (2005).

\bibitem{Tra11} T. A. Trainor and R. L. Ray, Phys.\ Rev.\ C {\bf 84}, 034906 (2011).

\bibitem{Ray11} R.~L.~Ray, Phys.\ Rev.\ D {\bf 84}, 034020 (2011);  T.~A.~Trainor and
                D.~J.~Prindle, {\it Improved isolation of the pp underlying event
                based on minimum-bias trigger-associated hadron correlations},
                arXiv:1310.0408 [hep-ph].

\bibitem{TraKet11}
T. A. Trainor and D. T. Kettler,
Phys. Rev.  {\bf C84}, 024910 (2011),  arXiv:1010.3048.

\bibitem{Mak15}
M.  Maksiak  (NA61/SHINE Collaboration), arXiv:1503.02470.

\bibitem{Gaz15}
M. Gazdzicki (NA61/SHINE Collaboration), EPJ Web Conf. {\bf 95},  01005  (2015), arxiv:1412.4243.

\bibitem{Lar15}
D. T. Larsen (NA61/SHINE Collaboration), Proceedings of International Conference on Strange Quark Matter,  Dubna, July 6-11, 2015 (to be published). 


\bibitem{Ser15}
Andrey Seryakov  (NA61/SHINE Collaboration),
 Proceedings of International Conference on Strange Quark Matter,  Dubna, July 6-11, 2015 (to be published). 






\bibitem{Pes79}
 M. E. Peskin, Nucl. Phys. {\bf B156}, 365 (1979);
G. Bhanot and M. E. Peskin, Nucl. Phys. {\bf B156}, 391 (1979).

\bibitem{Ynd02}
F. J. Yndurain,
{\it Low Energy Pion Physics},
 arXiv:hep-ph/0212282.

\bibitem{Won04}
C. Y. Wong,
Phys. Rev. {\bf C69}, 055202 (2004).



\bibitem{Won01}
C. Y. Wong and H. W. Crater, Phys. Rev. {\bf  C63},  044907 (2001).


\bibitem{dirac}  P.A.M. Dirac, Canad. J. Math. {\bf 2}, 129 (1950); Proc.
Roy. Soc. Sect. A {\bf 246}, 326 (1958); {\it Lectures on Quantum Mechanics}
(Yeshiva University, Hew York, 1964).


\bibitem{cra82}  P. Van Alstine and H.W. Crater, J. Math. Phys. {\bf 23},
1997 (1982); H. W. Crater and P. Van Alstine, Ann. Phys. (N.Y.) {\bf 148},
57 (1983). 

\bibitem{cra88}  H. W. Crater and P. Van Alstine, Phys. Rev. D1 {\bf 37},
1982 (1988).

\bibitem{saz89}  H. Sazdjian, Ann. Phys. {\bf 191, }52 (1989); Phys. Rev. D 
{\bf 33}, 3401 (1986).



\bibitem{cra92}  H. W. Crater, R. Becker, Cheuk-Yin Wong, and P. Van
Alstine, Phy. Rev. {\bf D46}, 5117 (1992).

\bibitem{cra96}  H. W. Crater, C. W. Wong, and C. Y. Wong, Intl. J. Mod.
Phys.-E {\bf 5}, 589 (1996); J. Mourad and H. Sazdjian, J. Phys. G {\bf 21}
267, (1995).


\bibitem{Cra04}
H. W. Crater, P. Van Alstine
 Phys.  Rev. {\bf D70}, 034026  (2004).

\bibitem{Cra09}
H. W. Crater, J. H. Yoon, and C. Y. Wong 
Phys. Rev. {\bf D79}, 034011 (2009).

\bibitem{Bar92}
T. Barnes and E.  S. Swanson, Phys. Rev.  {\bf D46}, 131 (1992).

\bibitem{Won02}
C. Y. Wong,  E. S. Swanson, and T. Barnes, 
Phys. Rev. {\bf C65}, 014903 (2002).

\bibitem{Adakpi04}
J. Adams et al. (STAR Collaboration)
Phys. Rev. Lett. 92, 112301 (2004).
% gives pi+=pi-,K+=K- data at 200 GeV


\bibitem{Wol11}
G. Wolschin, 
Europhys. Lett.  {\bf 95}, 61001 (2011),  (arxiv:1106.3636).

\bibitem{PHO11}
B. Alver,(PHOBOS Collaboration), Phys. Rev.  {\bf C83}, 024913 (2011).



\bibitem{Vov14}
V. Yu. Vovchenko, D. V. Anchishkin, and M. I. Gorenstein, Phys. Rev.  {\bf C 90}, 024916 (2014);
V. Uzhinsky , arXiv:1404.2026.

\bibitem{Wer06}
K. Werner, F. M. Liu and T. Pierog, Phys. Rev. {\bf C 74},  044902 (2006). [hep-ph/0506232].


\end{thebibliography}
\end{document}